\documentclass[12pt,a4paper]{article}
\usepackage{ragged2e}
\usepackage{subcaption} 
\usepackage{amsfonts}
\usepackage{amsmath}
\usepackage{array}
\usepackage{rotating}
\usepackage{amssymb}
\usepackage{bm}
\usepackage{geometry}
\usepackage{setspace}
\usepackage{graphicx}
\usepackage{lscape}
\usepackage{verbatim}
\usepackage{natbib}
\usepackage{xcolor}
\usepackage{calrsfs}
\usepackage{mathrsfs}
\usepackage[ruled]{algorithm2e}
\usepackage{algpseudocode}
\usepackage{tablefootnote}
\usepackage{tabularx}
\usepackage{threeparttablex}
\usepackage{longtable}
\usepackage{booktabs}
\usepackage{ragged2e}
\usepackage{enumitem}
\definecolor{winered}{rgb}{0.5,0,0}
\usepackage[bookmarks=true, bookmarksnumbered=true, allbordercolors={1 1 1}]{hyperref}
\hypersetup{
  colorlinks   = true, 
  urlcolor     = blue, 
  linkcolor    = blue, 
  citecolor    = winered,
}
\usepackage[open=true, numbered=true]{bookmark}
\usepackage{float}
\usepackage{fancyhdr}
\usepackage[toc,page]{appendix}
\usepackage{scalefnt}
\usepackage{afterpage}
\usepackage[left]{lineno}
\usepackage{caption}
\usepackage{varioref}
\usepackage{authblk}
\usepackage{cleveref}
\usepackage{amsthm}
\usepackage{multirow}

\DeclareFontFamily{OT1}{pzc}{}
\DeclareFontShape{OT1}{pzc}{m}{it}{<-> s * [0.900] pzcmi7t}{}
\DeclareMathAlphabet{\mathpzc}{OT1}{pzc}{m}{it}

\makeatletter
\def\blfootnote{\xdef\@thefnmark{}\@footnotetext}
\makeatother

\emergencystretch=\maxdimen
\theoremstyle{definition}

\usepackage{thmbox} 

\usepackage{calrsfs}
\DeclareMathAlphabet{\pazocal}{OMS}{zplm}{m}{n}

\def\wt{\widetilde}
\def\wh{\widehat}

\newcommand{\bB}{{\mathbf B}}
\newcommand{\bF}{{\mathbf F}}

\newcommand{\bV}{{\mathbf V}}

\newcommand{\bX}{{\mathbf X}}

\newcommand{\bff}{{\mathbf f}}

\newcommand{\bx}{{\mathbf x}}
\newcommand{\by}{{\mathbf y}}

\newcommand{\blambda}{\boldsymbol{\lambda}}

\newcommand{\bSigma}{\boldsymbol{\Sigma}}

\newcommand{\btheta} {\boldsymbol{\theta}}

\newcommand{\bLambda} {\boldsymbol{\Lambda}}

\newcommand{\bD}{{\mathbf D}}

\newcommand{\ve}{{\varepsilon}}
\renewcommand{\epsilon}{{\ve}}
\renewcommand{\hat}{\widehat}

\def\wt{\widetilde}
\renewcommand{\tilde}{\wt}

\begin{document}

\title{\bf\Large{A Supervised Screening and Regularized Factor-Based Method for Time Series Forecasting} }
\author{
Sihan Tu$^1$ and Zhaoxing Gao$^2$\thanks{\noindent Corresponding author: \texttt{zhaoxing.gao@uestc.edu.cn} (Z. Gao),  School of Mathematical Sciences, University of Electronic Science and Technology of China, Chengdu, 611731 P.R. China. We gratefully acknowledge the research support from the National Natural Science Foundation
of China (NSFC) under the grant numbers 12201558 and U23A2064.}\\ 
$^1$School of Management, Zhejiang University\\ $^2$School of Mathematical Sciences, University of Electronic Science and Technology of China
}

\date{}

\maketitle
\begin{abstract} 

Factor-based forecasting using Principal Component Analysis (PCA) is an effective machine learning tool for dimension reduction with many applications in statistics,  economics, and finance. This paper introduces a Supervised Screening and Regularized Factor-based (SSRF) framework that systematically addresses high-dimensional predictor sets through a structured four-step procedure integrating both static and dynamic forecasting mechanisms. 
The static approach selects predictors via marginal correlation screening and scales them using univariate predictive slopes, while the dynamic method screens and scales predictors based on time series regression incorporating lagged predictors. PCA then extracts latent factors from scaled predictors, followed by LASSO regularization to refine predictive accuracy. In the simulation study, we validate the effectiveness of SSRF and identify its 
parameter adjustment strategies in high-dimensional data settings. An empirical analysis of macroeconomic indices in China demonstrates that the SSRF method generally outperforms several commonly used forecasting techniques in out-of-sample predictions.

\noindent 

\bigskip

\noindent \emph{Keywords:} Factor-based Forecasting, Screening, Principal Component Analysis, High-Dimension, LASSO
\medskip

\noindent\textbf{JEL classification:} C33, C38 
\bigskip \medskip

\end{abstract}
\thispagestyle{empty} 

\newpage
\onehalfspace
\setcounter{page}{1}

\section{Introduction}


Advances in information technology have made high-dimensional data ubiquitous across various disciplines. For instance, investors now have access to thousands of stock options across hundreds of global markets, enabling more comprehensive yet complex decision-making processes. Extensive financial datasets, encompassing economic, trade, and investment activities, are constructed to uncover patterns influencing fluctuations in financial indices. While  the abundance of predictive variables offers a wealth of information, it also poses significant challenges, such as computational inefficiencies, increased costs, and the instability of the sample covariance matrix  (\cite{ledoit2004well}; \cite{bickel2008regularized}). Consequently, dimensionality reduction techniques and feature selection methods have emerged as powerful strategies to address these challenges.
Principal Component Analysis (PCA) is widely adopted for dimensionality reduction due to its effectiveness in condensing large datasets. However, as an unsupervised learning technique, PCA selects independent factors based solely on variance, which may not directly relate to the target variables of interest. Alternatively, penalized likelihood methods (PLMs) have risen to prominence for their ability to incorporate prior knowledge and constraints into the modeling process. Examples include the bridge regression introduced in \cite{frank1993statistical} and the Least Absolute Shrinkage and Selection Operator (LASSO) in \cite{tibshirani1996regression}. These techniques have been followed by innovations like the smoothly clipped absolute deviation (SCAD) method by \cite{fan2001variable}, and the elastic net approach by \cite{zou2005regularization}, among many others. Each of these methods brings a unique perspective to the challenge of variable selection, aiming to enhance model performance by mitigating overfitting and improving model interpretability. Despite their effectiveness, these methods can struggle with increased computational complexity, difficulties in parameter tuning, and the curse of dimensionality as the dimensions become large.

Factor models, pioneered by \cite{stock2002forecasting,stock2002macroeconomic}, have become fundamental tools for analyzing high-dimensional economic time series through latent factor extraction.  
Although conventional PCA delivers consistent factor estimation under the  foundational assumptions formulated by \cite{bai2002determining} \cite{fan2013thresh}, it remains sensitive to weak factor contamination in high-dimensional systems, potentially introducing systematic estimation bias, as showed in \cite{onatski2012asymptotics}, \cite{bai2023approximate} and \cite{huang2024time}.
Recent supervised approaches, exemplified by the predictive relevance-weighted scaling in \cite{scaledpca} and the dynamic adaptive reweighting regularization method in \cite{gao2024supervised}, have advanced dimension reduction by incorporating response guidance. 
However, existing literature has not fully examined model behavior in high-dimensional systems with latent factors exhibiting heterogeneous strength profiles. Additionally, fixed scaling mechanisms risk over-suppressing weakly influential factors while amplifying noise from spurious variable correlations.

In this paper, we propose a novel Supervised Screening and Regularized Factor-based (SSRF) model for time series forecasting, using a structured four-step procedure. Both static and dynamic forecasting approaches are employed.  For static forecasting, variables are selected based on strong marginal correlations with the target and scaled using predictive slopes from univariate regressions. For dynamic forecasting, variables are screened by $R^2$ values from time series regressions with lagged predictors and scaled by the corresponding regression coefficients.
After screening and scaling, PCA is applied to extract dominant factors, with three strategies are proposed to capture the primary variance structure for prediction. Finally, penalized regression techniques, such as LASSO, are used to refine the model, mitigating overfitting and enhancing predictive accuracy in high-dimensional settings.


To rigorously assess the performance of the proposed method, we conduct Monte Carlo simulations under various data-generating processes (DGPs) that mimic both static and dynamic forecasting scenarios. The simulations reveal that SSRF achieves robust factor recovery and forecasting accuracy, particularly in settings with strong predictive factors. Stricter screening thresholds enhance performance in high signal-to-noise environments, while relaxed thresholds preserve informational content when facing weak predictive factors coexisting with strong panel confounders or high noise levels. Compared to alternative methods such as PCA, Lasso, and Random Forest, SSRF consistently delivers lower mean squared forecast errors (MSFE), underscoring its effectiveness in isolating relevant signals and adapting to diverse forecasting challenges. 

Empirically, we forecast four Chinese macroeconomic indices: excess stock returns, stock volatility, the Consumer Price Index (CPI), and the Producer Price Index (PPI), using a comprehensive dataset of 54 key macroeconomic indicators in the Chinese economy, spanning from November 1996 to December 2019. These variables have been recognized in prior literature as crucial components for effective macroeconomic forecasting (\cite{mccracken2016fred}, \cite{tu2023}, and \cite{gao2024supervised}). We compare our method with the traditional PCA, the scaled PCA in \cite{scaledpca}, and Screened PCA  to highlight its  advantages  in factor extraction. Additionally, we benchmark SSRF against linear regression, SCAD (\cite{fan2001variable}), and elastic net (\cite{zou2005regularization}) to assess its forecasting performance. The results show that SSRF consistently outperforms these commonly used forecasting models in out-of-sample prediction.

Our contributions to the field are threefold. First, we introduce a novel Supervised Screening and Regularized Factor-based (SSRF) forecasting method, which integrates both static and dynamic techniques. This four-step approach maximizes the information exchange between target variables and predictors, offering improvements over previous methodologies. Unlike \cite{bai2008forecasting}, who apply Lasso for variable selection on raw data, our method applies the selection process to the factors derived from PCA, ensuring a more refined model. Additionally, while \cite{goulet2022machine} provide a comprehensive comparison of nonlinear and regularized machine learning methods, including PCA and PLMs, our work extends this by offering a detailed evaluation of the proposed method under varying factor and noise intensities in a simulation study. This allows us to identify scenarios where the method excels and highlight its potential limitations.

Second, beyond the traditional integration of factor analysis with Lasso, we incorporate supervised static and dynamic factors in an empirical application. This dual-dimensional approach enables Lasso to select variables across both temporal and feature dimensions, ensuring the identification of the most informative factors while mitigating the risk of information loss inherent in multi-stage dimensionality reduction. This innovation enhances the robustness and adaptability of the model in high-dimensional settings.

Third, unlike most research that centers on the U.S. economy, our study focuses on Chinese macroeconomic conditions and their relationship with the stock market. This focus is particularly important given the stricter financial regulations and market constraints in China compared to the U.S., providing unique insights into the applicability of advanced forecasting methods in emerging markets with distinct economic structures.


The remainder of this article is organized as follows: Section~\ref{sec2} provides an in-depth introduction to our proposed model and methodology. Section~\ref{sec7} studies the finite-sample performance of the proposed approach via simulation. Section~\ref{sec3} presents the macroeconomic dataset used in the empirical prediction task. Section~\ref{sec4} examines the in-sample forecasting performance of the proposed method, and Section~\ref{sec5} assesses the out-of-sample performance. Finally, Section~\ref{sec6} concludes the study.

\section{Methodology}\label{sec2}
Let $\bX = (\bx_1, \ldots, \bx_p)' = (\bX_1, \ldots, \bX_{T})$ denote a \(p \times T\) matrix of observed predictors or features, where $p$ and $T$ denote the number of predictors and the number of observations, respectively. The superscript $'$ denotes the transpose of a
vector or matrix. For $1\leq t\leq T$ and $1\leq j\leq p$, $\bX_t=(x_{1,t},...,x_{p,t})'$ signifies a $p$-dimensional feature vector and $\bx_j = (x_{j,1},\ldots,x_{j,T})'$ represents a vector consisting of the $T$ observations for the $j$-th predictor.
Furthermore, let $\by = (y_{1}, \ldots, y_{T})'$ be a \(T\)-dimensional observable vector of a scalar target variable. For $h>0$, our objective is to estimate the latent factor structure of the high-dimensional data that captures the relationship between $\bX_t$ and the target variable $y_{t+h}$, and subsequently use it to predict the target variable $y_{t+h}$. Each sequence of the data has been standardized to have zero mean and unit variance. Our methodology relies on the principle that only a proportion of all the predictors have significant forecast ability for the target variable of interest. Next, we  present the static and dynamic modeling approaches of the SSRF model, along with the strategies for parameter tuning.

\subsection{Static Forecasting}
Static forecasting refers to factor-based forecasting using contemporaneous factors extracted from a panel of predictors. The procedure consists of the following four steps:

\begin{enumerate}
\item \textbf{Correlation Learning and Screening.}
Our approach begins with a filtering step to identify predictors that show a strong relationship with the target variable. Following the sure independence screening framework(SIS) outlined by \cite{fan2008sure}, we assess the marginal correlation of each predictor with the response variable. We introduce a $p$-dimensional vector of sample correlations $\mathbf{\omega} = (\omega_1, \ldots, \omega_p)'$, where $\{\omega_j\}_{j=1}^p$ are calculated using the following formula:
\begin{equation}
    \omega_{j}=|\frac{1}{T}\sum_{t=1}^{T-h}x_{j,t}y_{t+h}|, \quad \text{for} \quad j = 1, 2, \ldots, p.
\end{equation}
For a predetermined threshold $ \kappa_1 $, the subset of covariates that surpass this threshold is denoted as $ M_{\kappa,1} $, where
\begin{equation}
    M_{\kappa,1} = \{1 \leq j \leq p : \omega_j \geq \kappa_1\}.
\end{equation}
If $M_{\kappa,1}$ comprises $k$ predictors, then the corresponding sub-feature space can be represented as $ \tilde{\bX} = (\tilde{\bx}_{1}, \ldots, \tilde{\bx}_{k})'$, where $\{\tilde{\bx}_{i}\}_{i=1}^k$ are the selected $k$ features of length $T$.


\item \textbf{Scaling Predictors.}
After screening the covariates, we integrate the response and sub-sample spaces using the  scaled PCA introduced by \cite{scaledpca}. We begin by regressing the target $y_{t+h}$ on each of the selected  predictors, i.e.,
\begin{equation}
    y_{t+h} = \varphi_j\tilde{x}_{j,t} + u_{j,t+h},\quad\mathrm{for}\quad j=1,2,\ldots,k, t=1,...,T-h.
\end{equation}
Here, $\varphi_j$ represents the regression coefficient for the \(j\)-th predictor, and $u_{j,t+h}$ is the error term associated with each regression. Let $\wh\varphi_j$ be the OLS estimator, then $\{\hat{\varphi}_j \tilde{\bx}_{j}\}_{j=1}^k$ are the scaled predictors. 
This scaling emphasizes that different  predictors  have varying predictive power. The resulting scaled predictors can be represented as $\dot{\bX} = (\hat{\varphi}_1 \tilde{\bx}_1, \ldots, \hat{\varphi}_k \tilde{\bx}_k)' = (\dot{\bx}_1,\dots,\dot{\bx}_k)'$, where $\dot{\bX}$ is a $k \times T$ matrix.

\item \textbf{Factor Modeling.}
To extract the latent factor structures, we perform Principal Component Analysis (PCA) on the set of scaled predictors $\dot{\bX}$. 
Let  $\bff_{t} \in \mathbb{R}^{r}$ be an $r$-dimensional latent factor process,  and $\blambda_j \in \mathbb{R}^{r}$ be the loading vector for the \(j\)-th scaled predictor $x_{j,t}$. 
We denote the matrix of latent factors as \(\bF = (\bff_{1}, \ldots, \bff_{T})'\in R^{T\times r}\)and the matrix of loading vectors as \(\bLambda = (\blambda_1, \ldots, \blambda_k)'\in R^{k\times r}\).
The objective of  PCA is to minimize the following expression,
\begin{equation}
(\widehat{\bF},\widehat{\bLambda})=\arg\min_{\bF,\bLambda} \frac{1}{Tk}\|\dot{\bX}-\bLambda \bF'\|_{F}^{2}, \quad \text{subject to}\quad \frac{1}{T} {\bF}'{\bF}=\mathbf{I}_{r},
\label{pcamse}
\end{equation}
where \(\mathbf{I}_r\) is an \(r \times r\) identity matrix. 
The columns of \(\widehat{\bF} / \sqrt{T}\) are the eigenvectors corresponding to the \(r\) largest eigenvalues of the \(T \times T\) matrix \(\tilde{\bSigma} = \frac{1}{k} \dot{\bX}' \dot{\bX}\). The loading matrix is derived as \(\widehat{\bLambda} = T^{-1} \dot{\bX} \widehat{\bF}\).

\item \textbf{Lasso Selection and Prediction.} 
While PCA reduces dimensionality and captures the underlying structure of the data, Lasso fine-tunes the model by selecting a subset of predictors that are most relevant for the target variables. We consider the linear regression model:
\begin{equation}
    y_{t+h} = \btheta' \wh\bff_t + \varepsilon_{t+h}, \quad\mathrm{for}\quad t=1,...,T-h.
\end{equation}
where $\btheta \in \mathbb{R}^{r}$ is a vector of regression coefficients for the latent factor vector $\wh\bff_t$, extracted from Step 3, and $\varepsilon_{t+h}$ is the error term at time $t+h$. Then, we employ Lasso for variable selection, which is formulated as:
\begin{equation}
\wh\btheta_{lasso} = \arg\min_{\btheta\in R^r} \left\{ \frac{1}{T} \sum_{t=1}^{T-h} \| y_{t+h} - \btheta' \hat{\bff}_t \|^2 + \psi \|\btheta\|_1 \right\},
\end{equation}
where $\psi$ is the regularization parameter that controls the penalty’s strength on the absolute size of the regression coefficients.  Finally, the prediction of $y_{T+h}$ is given by $\wh y_{T+h}=\wh \btheta'\wh\bff_T$.

\end{enumerate} 

\subsection{Dynamic Forecasting}

Dynamic forecasting refers to factor-based forecasting that utilizes both contemporaneous factors and their lagged values to capture temporal dependencies in the data.
The procedure differs from static forecasting in that the first two steps are replaced with the following:

\begin{enumerate}[label={\arabic*.$'$}]
\item \textbf{Dynamic Screening.} Unlike the Step 1 above, which focuses on the marginal correlation between the contemporaneous predictor and the target one, we incorporate the  lagged predictors and make use of time series regression as follows:
\begin{equation}\label{eqa:dym}
y_{t+h}\approx\widehat{\mu}_i+\widehat{\gamma}_{i,0}x_{i,t}+\widehat{\gamma}_{i,1}x_{i,t-1}+...+\widehat{\gamma}_{i,q_i-1}x_{i,t-q_i+1},~t=q_i,...,T-h,
\end{equation}
where $\widehat{\gamma}_{i,t-q_i+1}$ represents the regression coefficients corresponding to the $q_i-1$ lagged variables of the $i$-th predictor. A screening strategy is then applied based on the $R^2$ value of the regression model in Eq. (\ref{eqa:dym}) for each predictor, excluding those with low $R^2$. The subset of covariates that meet the $R^2$ threshold is denoted as $M_{\kappa,2}$, where
\begin{equation}
M_{\kappa,2} = \{1 \leq j \leq p : R^2_j \geq \kappa_2\}.
\end{equation}
\item \textbf{Dynamic Scaling} This approach is motivated by the intrinsic dependencies within time series data, where more recent observations tend to carry greater informational content and are more effective in capturing temporal dynamics.  Our methodology aligns with the research presented in \cite{gao2024supervised} for scaling time series data. Each predictor selected by Step $1'$ can be scaled as:
$\widehat{x}_{i,t}=\widehat{\gamma}_{i,0}x_{i,t}+\widehat{\gamma}_{i,1}x_{i,t-1}+...+\widehat{\gamma}_{i,q_i-1}x_{i,t-q_i+1}$. Consequently,
the corresponding dynamic scaling sub-feature space is represented by $\ddot{\bX}$.

\end{enumerate}

To summarize, factors can be extracted using two methods:
\begin{enumerate}
    \item a static approach (Steps $1+2+3$), referred to as feature-screened and scaled PCA (SSPCA-f), which generates the factor space $\bF_{f}$.
    \item a dynamic approach  (Steps $1'+2'+3$), referred to as dynamic-screened and scaled PCA (SSPCA-d), which yields the factor space $\bF_{d}$.
\end{enumerate}
These approaches correspond to three variants of SSRF in Step 4: (1) using $\bF_{f}$, (2) using $\bF_{d}$, or (3) using a hybrid factor space, $\bF_{hyb}$, which combines columns from both $\bF_{f}$ and $\bF_{d}$, denoted as SSRF(1), SSRF(2), and SSRF(3), respectively. One motivation for adopting the third factor modeling strategy is that, while both dimensions originate from the same underlying dataset, they each provide valuable insights. By integrating Lasso regression after PCA, we can achieve more precise variable selection, mitigating redundancy and enhancing model prediction.

\subsection{Choice of Tuning Parameters}
Our framework involves the following key tuning parameters:
\begin{enumerate}
    \item \textbf{Screening Thresholds ($\kappa_1,\kappa_2$)}: In contrast to scaling methods that preserve the original space structure through linear transformations, the screening mechanism achieves an essential reconstruction of data structure via feature subset selection, where thresholding is crucial. Commonly employed methods include SIS proposed by \cite{fan2008sure}, which retains a predefined number of predictors, such as $[n/logn]$, where $n$ refers to the sample size, and the automatic thresholding method based on t-statistics introduced by \cite{bai2008forecasting}. Optimization-driven approaches, such as the Orthogonal Greedy Algorithm (OGA), enable precise feature selection through iterative computation but at the cost of higher computational complexity.  Considering the variability in predictive signals and noise inherent in collected data, we employ an expanding time window cross-validation in our experiments to determine the optimal number of retained variables through a grid search over the parameter space $ \{0.1, 0.2, 0.5,0.75,1\}$. In our simulation studies, we provide recommended screening strategy under different factor strength and noise levels, demonstrating that stricter thresholds enhance performance in high signal-to-noise ratio environments, while more inclusive thresholds mitigate information loss in scenarios with weaker predictive factors.
    \item \textbf{Lag Order ($q_i$)}: $q_i$ is selected for each predictor using an information criterion, such as the Akaike Information Criterion (AIC). Empirical analysis suggests that $q_i \in \{1,2\}$ is generally sufficient for monthly macroeconomic data.
    \item \textbf{Number of Factors ($r$)}: Determining the optimal number of factors in factor analysis is pivotal for revealing the inherent structure within the dataset. This number can be estimated using the information criterion in \cite{bai2002determining} or the ratio-based method in \cite{lam2012factor}. Here, drawing from the work detailed in \cite{lam2012factor}, we use a ratio-based estimator to determine the number of factor as follows,
    \begin{equation}
    \widehat{r}=\arg\min_{1\leq i\leq R}  \widehat{\lambda}_{i+1}/\widehat{\lambda}_i, 
    \end{equation}
    where $\widehat{\lambda}_1\geq\cdots\geq\widehat{\lambda}_p$ are the eigenvalues of a covariance matrix of the constructed panel of predictors, and $0<R<p$ is a prescribed constant, which can be taken as $R=[p/2]$.
    Alternatively, we may keep a sufficiently large number of factors and apply the Lasso  to select the factors that have significant predictive power for the target variables of interest. 
    \item \textbf{Lasso Penalty ($\psi$)}: A larger $\psi$ increases the penalty, leading to more coefficients being shrunk to zero, thus promoting sparsity. The optimal value of $\psi$ is typically estimated using cross-validation to achieve the best predictive performance.
\end{enumerate}

\section{Numerical Study}\label{sec7}
In this section, we conduct Monte Carlo simulations to evaluate the finite-sample performance of the proposed method and examine its strengths and limitations under various conditions. 
We follow the data-generating process (DGP) in \cite{bai2023approximate}, where the observed $p\times T$ data matrix $\bX$ is generated element-wise as $x_{jt} = \blambda_j'\bff_t+ \sigma_je_{jt}$.  The latent factor structure comprises  $\bff_t \sim N(0,I_r)$ and $\blambda_j \sim N(0,I_r)\bD\bB/\sqrt{p}$, with diagonal matrices $\bD$ and $\bB$ encoding factor strengths. Specifically, each diagonal entry of $\bB$ is defined as $\bB_{jj} = p^{\alpha_j / 2}$ for $j = 1, 2, \dots, r,$ where the exponents $\alpha_j \in [0,1]$ quantify the relative importance of each factor. Larger values of $\alpha_j$ indicate stronger factors, while smaller values correspond to weaker factors.
To introduce idiosyncratic variability, noise terms $e_{jt} \sim N(0,1)$ are scaled by variances $\sigma_j$, which are independently sampled from a uniform distribution. This setup ensures controlled variability in noise magnitudes across observations. We consider three distinct methods for generating the target variable $y_{t+h}$, with $h=1$, $p=500$ and $T=100$:
\begin{itemize}
    \item  \textbf{DGP1:} The target variable is defined as $y_{t+h} = \btheta'\bff_t + \epsilon_{t+h}$ with $r = 4$, $\epsilon_{t+h} \sim N(0,1)$, $\btheta = (1, 0, 2, 5)'$, $\bD^2 = diag(3,2,1,0.7)$ and $\sigma_j \sim U(0.1,0.5)$. 
    \item \textbf{DGP2:} The target variable is similarly defined as \textbf{DGP1}, but with $\btheta = (0,0,0,5)'$ and $\sigma_j \sim U(0.1,0.5)$ while maintaining $r=4$, $\bD^2 = diag(3,2,1,0.7)$ and  $\epsilon_{t+h} \sim N(0,1)$ .
    \item \textbf{DGP3:} The target variable incorporates lagged factors, $y_{t+h} = \btheta'(\bff_t,\bff_{t-1}) + \epsilon_{t+h}$ with $r = 2$,  $\epsilon_{t+h} \sim N(0,1)$, $\btheta = (1,0,0.7,0)'$ , $\bD^2 = diag(1,1,1,1)$ and $\sigma_j \sim U(0.8,1)$.
\end{itemize}
DGP1 and DGP2 utilize contemporaneous factors, whereas DGP3 introduces lagged factors into the mix. In both DGP1 and DGP3, multiple factors are predictive, with the second factor deliberately set to be non-predictive. In contrast, DGP2 represents an extreme scenario where solely the fourth factor is predictive.
Since the estimated factors  $\wh\bff_t$ are only identifiable up to a rotation within the factor model framework, we conduct Lasso regression on the transformed factors $\bV \wh\bff_t$. Here, $\bV$ denotes the right singular matrix obtained from the singular value decomposition (SVD) of the scaled loading matrix. To rigorously assess the model's efficacy in factor recovery, two key metrics are employed: the frequency with which nonzero coefficients are accurately identified, and the precision of factor recovery, as gauged by the norm  $\|\wh\bF\wh\bF'- \bF\bF'\|_2$.

Tables \ref{tab:statcover} and Table \ref{tab:dynacover} summarize the factor recovery performance for DGP1, DGP2, and DGP3 under varying factor strengths.  Stronger factors (e.g., $(1,1,1,1)$) provide a pronounced signal-to-noise ratio, while weaker settings (e.g., $(0.5,0.5,0.5,0.5)$) dilute the overall information strength. Comparisons, such as between (1,1,1,1) and (1,0.5,1,1), illustrate shifts in relative factor strength, where non-predictive factors are less dominant, and vice versa.
For DGP1, settings with strong predictive factors, (e.g., (1,1,1,1) and (1,0.5,1,1)) benefit from stricter screening threshold (e.g., $k/p$ = 0.1), achieving higher recovery rates despite  a slight trade-off in accuracy due to aggressive filtering. This highlights the ability of stringent thresholds to effectively isolate relevant components and suppress noise in high-information environments. As the overall factor strength weakens (e.g., (0.5,0.5,0.5,0.5)), higher screening thresholds continue to improve recovery rates, but the increased impact of noise reduces overall accuracy.
However, when irrelevant factors dominate, as in (0.5,1,0.5,0.5) for DGP1 or (1,1,1,0.5) for DGP2, stricter screening may inadvertently exclude weak but relevant signals, lowering recovery rates. In such low signal-to-noise conditions, a more lenient threshold helps retain critical predictive information, reducing the risk of omitting key components. Similar results can be observed for DGP3, as shown in Table~\ref{tab:dynacover}.

\begin{table}[!ht]
\centering
\renewcommand\arraystretch{1.3}
\caption{Recovery rate  and accuracy evaluation of recovered latent factor across different cases and strength configurations for DGP1 and DGP2}
\label{tab:statcover}
\resizebox{\textwidth}{!}{
\begin{threeparttable}
    \begin{tabular}{ccccccccccccc}
\hline
\multicolumn{2}{c}{\textbf{DGP1}} & \multicolumn{2}{c}{(1,1,1,1)} &  & \multicolumn{2}{c}{(1,0.5,1,1)}          &  & \multicolumn{2}{c}{(0.5,1,0.5,0.5)} &  & \multicolumn{2}{c}{(0.5,0.5,0.5,0.5)} \\ \cline{1-4} \cline{6-7} \cline{9-10} \cline{12-13} 
$k/p$       &        & recovery rate           & $\|\wh\bF\wh\bF' - \bF\bF'\|_2 $            &  & recovery rate              & $\|\wh\bF\wh\bF' - \bF\bF'\|_2$                &  & recovery rate              & $\|\wh\bF\wh\bF' - \bF\bF'\|_2$                &  & recovery rate               & $\|\wh\bF\wh\bF' - \bF\bF'\|_2$                 \\ \hline
1                &        & 0.55              & 0.727             &  & 0.96                 & 0.309                &  & 0.28                 & 0.861                &  & 0.46                  & 0.725                 \\
0.75             &        & 0.54              & \textbf{0.715}    &  & 0.97                 & \textbf{0.303}                &  & 0.30                 & 0.848                &  & 0.41                  & 0.742                 \\
0.5              &        & 0.61              & 0.717             &  & 0.96                 & 0.314                &  & \textbf{0.32}        & \textbf{0.829}       &  & 0.56                  & \textbf{0.715}        \\
0.2              &        & 0.77              & 0.767             &  & \textbf{0.99}                 & 0.306       &  & 0.12                 & 0.897                &  & 0.64                  & 0.738                 \\
0.1              &        & \textbf{0.81}     & 0.783             &  & 0.98        & 0.360                &  & 0.17                 & 0.884                &  & \textbf{0.65}         & 0.805                 \\ \hline
\multicolumn{2}{c}{\textbf{DGP2}} & \multicolumn{2}{c}{(1,1,1,1)}          &  & \multicolumn{2}{c}{(0.5,0.5,0.5,1)} &  & \multicolumn{2}{c}{(1,1,1,0.5)}              &  & \multicolumn{2}{c}{(0.5,0.5,0.5,0.5)}          \\ \cline{1-4} \cline{6-7} \cline{9-10} \cline{12-13} 
$k/p$       &        & recovery rate           & $\|\wh\bF\wh\bF' - \bF\bF'\|_2$             &  & recovery rate              & $\|\wh\bF\wh\bF' - \bF\bF'\|_2$                &  & recovery rate              & $\|\wh\bF\wh\bF' - \bF\bF'\|_2$                &  & recovery rate               & $\|\wh\bF\wh\bF' - \bF\bF'\|_2$                 \\ \hline
1                &        & 0.40              & 1.116             &  & 1.00                 & 1.107                &  & \textbf{1.00}        & \textbf{1.147}       &  & 0.35                  & 1.106                 \\
0.75             &        & 0.49              & 1.116             &  & 1.00                 & 1.107                &  & 1.00                 & 1.148                &  & 0.40                  & 1.106                 \\
0.5              &        & 0.77              & 1.117             &  & 1.00                 & 1.107                &  & 1.00                 & 1.149                &  & 0.58                  & 1.121                 \\
0.2              &        & \textbf{0.99}     & \textbf{1.092}    &  & \textbf{1.00}        & \textbf{1.106}       &  & 1.00                 & 1.153                &  & 0.83                  & 1.105                 \\
0.1              &        & 0.99              & 1.093             &  & 1.00                 & 1.108                &  & 0.91                 & 1.149                &  & \textbf{0.97}         & \textbf{1.091}        \\ \hline
\end{tabular}%
\begin{tablenotes}
\item NOTE: This table summarizes the performance of factor recovery under different configurations of factor strengths and data generation processes, measured by the frequency of correctly identifying nonzero coefficients (recovery rate) and the accuracy of factor recovery ($\|\wh\bF\wh\bF' - \bF\bF'\|_2$).$k/p$ represents the percent of reserved variables. We use 100 replications for each configuration.
\end{tablenotes}
\end{threeparttable}%
}
\end{table}

\begin{table}[!ht]
\centering
\renewcommand\arraystretch{1.3}
\caption{Recovery rate  and accuracy evaluation of recovered latent factor across different configurations for DGP3}
\label{tab:dynacover}
\resizebox{\textwidth}{!}{
\begin{threeparttable}
    \begin{tabular}{ccccccccccccccc}
\hline
{\textbf{DGP3}} & \multicolumn{2}{c}{1,1}      &  & \multicolumn{2}{c}{\textbf{1,0.5}} &  & \multicolumn{2}{c}{0.5,1}       &  & \multicolumn{2}{c}{\textbf{0.2,1}} &  & \multicolumn{2}{c}{\textbf{0.2,0.2}} \\ \cline{1-3} \cline{5-6}\cline{8-9}\cline{11-12}\cline{14-15}
{$k/p$}     & recovery rate    & $\|\wh\bF\wh\bF' - \bF\bF'\|_2$          &  & recovery rate       & $\|\wh\bF\wh\bF' - \bF\bF'\|_2$             &  & recovery rate       & $\|\wh\bF\wh\bF' - \bF\bF'\|_2$          &  & recovery rate         & $\|\wh\bF\wh\bF' - \bF\bF'\|_2$           &  & recovery rate          & $\|\wh\bF\wh\bF' - \bF\bF'\|_2$            \\ \hline
1                      & 0.99       & \textbf{1.126} &  & 1             & 1.131             &  & 0.26          & 1.097          &  & \textbf{0.75}   & 1.134           &  & 0.96             & \textbf{1.129}   \\
0.75                   & 0.98       & 1.126          &  & 1             & 1.132             &  & 0.35          & 1.091          &  & 0.74            & 1.131           &  & 0.96             & 1.130            \\
0.5                    & 0.99       & 1.130          &  & 1             & 1.134             &  & 0.38          & \textbf{1.090} &  & 0.57            & 1.126           &  & \textbf{0.98}    & 1.132            \\
0.2                    & 1          & 1.131          &  & \textbf{1}    & 1.133             &  & \textbf{0.39} & 1.106          &  & 0.08            & 1.086           &  & 0.97    & 1.132            \\
0.1                    & \textbf{1} & 1.135          &  & \textbf{1}    & \textbf{1.128}    &  & 0.28          & 1.114          &  & 0.02            & \textbf{1.074}  &  & 0.9              & 1.134            \\ \hline
\end{tabular}%
\begin{tablenotes}
\item NOTE: This table summarizes the performance of factor recovery under different configurations of factor strengths, measured by the frequency of correctly identifying nonzero coefficients (recovery rate) and the accuracy of factor recovery ($\|\wh\bF\wh\bF' - \bF\bF'\|_2$). $k/p$ represents the percent of reserved variables. We use 100 replications for each configuration.
\end{tablenotes}
\end{threeparttable}%
}
\end{table}

To measure the model’s forecasting performance, we compute the mean squared forecast error (MSFE), expressed as:
\begin{equation}\label{msfe}
    MSFE = \sqrt{\frac{1}{T-t} \sum_t^T (y_{t+h}-\widehat{y}_{t+h})^{2}}.
\end{equation}
Here, $y_{t+h}$ denotes the $h$-step actual target variable, and $\hat{y}_{t+h}$ refers to its predicted value. Lower MSFE values indicate higher forecasting accuracy. We employ an expanding window approach for out-of-sample forecasting, utilizing all available data up to a time period $t$ to forecast the value for the subsequent time period $t+h$. In the simulation study, we set $t=80$ and $T=100$. 

We demonstrate the out-of-sample MSFE results for DGP1 and DGP3 across various factor strengths in Table~\ref{tab:simupre}, considering alternative methods such as sPCA, PLS, adaptive Lasso, and random forest. Under configurations with strong predictive factors, including (1,1,1,1) and (1,0.5,1,1), SSRF(1) consistently delivers the lowest MSFE. This is because the strict screening threshold effectively isolates the dominant predictive factors while suppressing noise, leading to improved forecasting accuracy. In contrast, weaker factor settings lead to higher forecasting errors, as the dominance of noise makes identifying valuable information more challenging.
For dynamic factor models in DGP3, SSRF(2) shows a distinct advantage, outperforming other methods across all configurations. This highlights the different focuses of SSRF(1) and SSRF(2) on factor manifestations. 
 As shown in the table, the proportion of retained factors is reported alongside the best-performing values, providing insight into the role of screening thresholds.  Consistent with earlier results: strict screening is effective when predictive factors dominate, while inclusive thresholds are preferable when irrelevant factors prevail to preserve critical information.
\begin{table}[!ht]
\centering
\renewcommand\arraystretch{1.3}
\caption{MSFE comparison for DGP1 and DGP3 across different models and factor strength configurations}
\label{tab:simupre}
\resizebox{\textwidth}{!}{
\begin{threeparttable}
    \begin{tabular}{cccccccccc}
\hline
                     & \multicolumn{4}{c}{\textbf{DGP1}}                                     &                      & \multicolumn{4}{c}{\textbf{DGP3}}                                     \\ \cline{2-5} \cline{7-10} 
                     & 1,1,1,1         & 1,0.5,1,1       & 0.5,1,0.5,0.5   & 0.5,0.5,0.5,0.5 &                      & 1,1             & 1,0.5           & 0.5,1           & 0.5,0.5         \\ \hline
SSRF(2)              & 0.942          & 0.918          & 1.146          & 1.007          &                      & \textbf{0.894} & \textbf{0.948} & \textbf{1.070} & \textbf{1.007} \\
SSRF(1)              & \textbf{0.910} & \textbf{0.891} & \textbf{0.931} & \textbf{0.969} & \textbf{}            & 1.288          & 1.262          & 1.160          & 1.209          \\
\multicolumn{1}{l}{} & (0.1)           & (0.1)           & (0.75)          & (0.2)           & \multicolumn{1}{l}{} & (0.1)           & (0.75)          & (0.75)          & (0.5)           \\
PCA                  & 1.089          & 1.178          & 1.060          & 1.039          &                      & 0.929          & 1.355          & 1.233          & 1.290          \\
sPCA                 & 0.944          & 0.916          & 0.932          & 0.980          &                      & 1.339          & 1.245          & 1.293          & 1.379          \\
dPCA                 & 0.971          & 0.909          & 1.113          & 1.040          &                      & 1.119          & 1.019          & 1.101          & 1.055          \\
Lasso                & 1.412          & 1.362          & 1.815          & 1.829          &                      & 1.191          & 1.143          & 1.194          & 1.154          \\
Adaptive Lasso       & 1.259          & 1.145          & 1.383          & 1.620          &                      & 0.991          & 1.346          & 1.307          & 1.358          \\
SCAD                 & 5.168          & 5.327          & 5.104          & 4.978          &                      & 1.232          & 1.522          & 1.614          & 1.519          \\
PLS                  & 1.092          & 1.097          & 1.244          & 1.175          &                      & 1.107          & 1.402          & 1.296          & 1.336          \\
ENet                 & 1.182          & 1.138          & 1.309          & 1.485          &                      & 0.949          & 1.409          & 1.303          & 1.361          \\
RandomForest         & 1.669          & 1.562          & 2.976          & 2.510          &                      & 0.964          & 1.362          & 1.508          & 1.429          \\
SVM                  & 1.539          & 1.740          & 2.294          & 1.946          &                      & 1.094          & 1.334          & 1.398          & 1.310          \\
GBM                  & 1.746          & 1.693          & 2.130          & 1.976          &                      & 1.008          & 1.394          & 1.396          & 1.615          \\ \hline
\end{tabular}%
\begin{tablenotes}
\item  NOTE: Different thresholds are applied to these configurations when using SSRF. We select the screening thresholds taht yield the model’s optimal performance, with the proportion of reserved variables ( $k/p$ ) indicated in parentheses.
\end{tablenotes}
\end{threeparttable}%
}
\end{table}

\section{Data}\label{sec3}
In our empirical study, we employ the proposed SSRF method to forecast the 1-month ahead excess stock return, stock volatility, Consumer Price Index (CPI), and Producer Price Index (PPI) using 54 macroeconomic variables. The excess stock return is measured by the monthly market value-weighted excess return of Chinese A-share listed companies, excluding “ST” and “PT” companies, with the one-year fixed deposit interest rate serving as the risk-free rate. Stock volatility is calculated as the square root of the \texttt{svar} values, which represent the sum of squared daily returns of the Shanghai Stock Exchange (SSE) Composite Index over a corresponding month.

Our data is sourced from the China Stock Market and Accounting Research (CSMAR) database (\url{https://data.csmar.com/}), the National Bureau of Statistics (\url{https://www.stats.gov.cn/}), and the Wind Economic Database (\url{https://www.wind.com.cn/}). Similar to \cite{mccracken2016fred}, \cite{tu2023}, and \cite{gao2024supervised}, these variables are categorized into seven key macroeconomic domains: Output (OUT), Stock Market (SM), Prices (PR), Interest and Exchange Rates (IER), Money and Credit (MC), Consumption (CON), and Investment (INV). 
Due to the ongoing refinement of China's systematic macroeconomic data compilation, the macroeconomic variables differ from those of the U.S. as outlined by \cite{mccracken2016fred}. Our dataset consists of 54 variables, with detailed descriptions and the transformation codes used to ensure stationarity for each macroeconomic variable provided in Table~\ref{tab:varidescribe} in the Appendix. The dataset spans the period from November 1996 to December 2019. However, data after 2019 were excluded due to the unique economic impact of the COVID-19 pandemic on China. The data from November 1996 to June 2013 are used as the training sample, while the remaining portion serves as the test sample.

\begin{figure}[ht]
  \includegraphics[width = \textwidth]{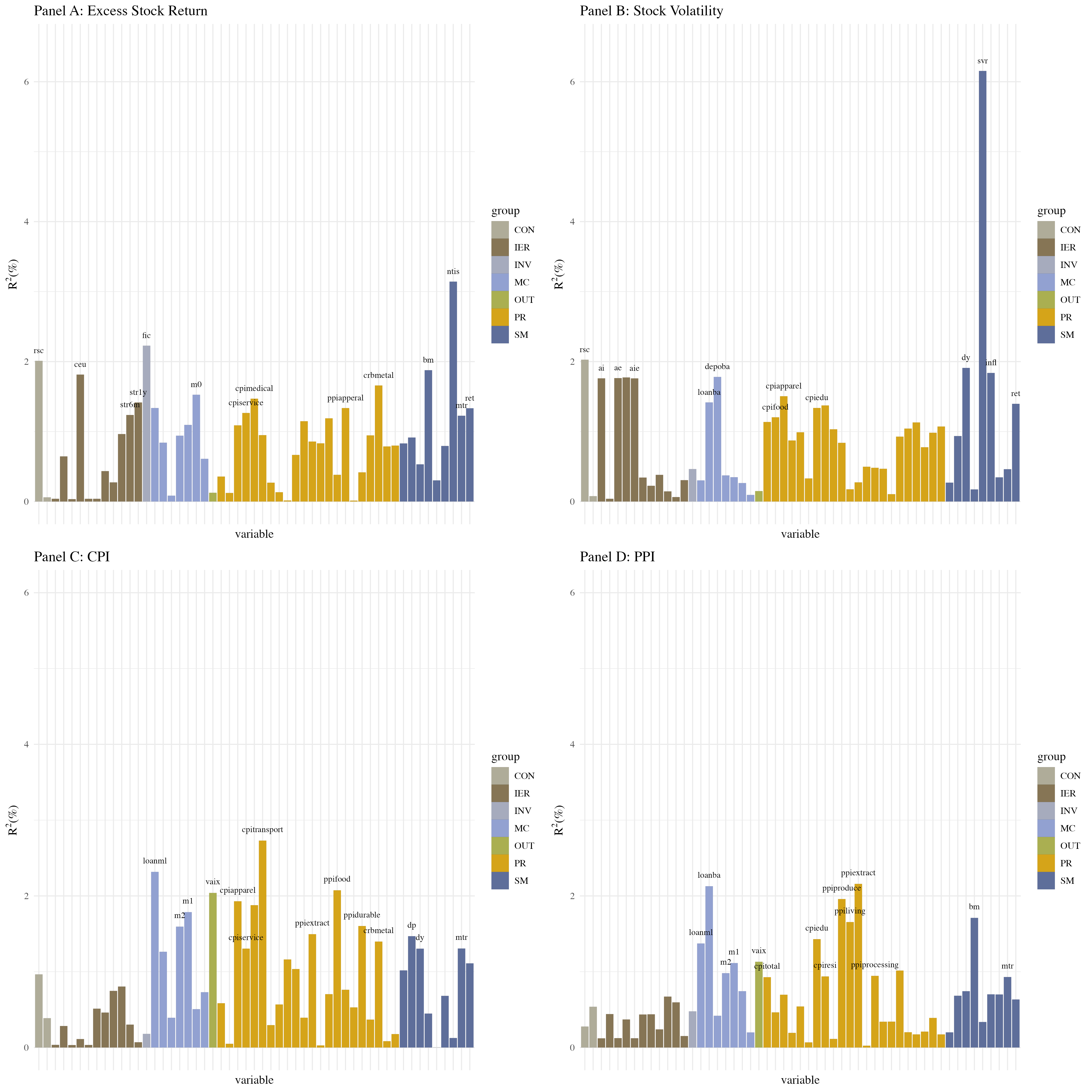}
  \caption{In-sample forecasting of macroeconomic variables. The figure presents the in-sample $R^2$ (in percentage) values for predicting 1-month ahead excess stock returns (Panel A), stock volatility (Panel B), CPI (Panel C), and PPI (Panel D) using 54 macroeconomic variables. These variables are grouped into seven categories: Output (OUT), Stock Market (SM), Prices (PR), Interest Rates and Exchange Rates (IER), Money and Credit (MC), Consumption (CON), and Investment (INV). The $R^2$ is defined as ${\sum_{i=1}^{T}(\hat{y}{i}-\overline{y})^{2}}/{\sum_{i=1}^{T}(y_{i}-\overline{y})^{2}}$, where $\hat{y}$ represents the fitted values and $\overline{y}$ is the in-sample mean. All data are publicly available. The sample period is from November 1996 to June 2013.}
  \label{fig:samr2}
\end{figure}

In the initial exploration, we investigate the predictive power of each individual predictor on the focal variables. \autoref{fig:samr2} presents the in-sample $R^2$ values for predicting the 1-month ahead excess stock return, change in stock volatility, change in CPI and change in PPI, utilizing each of the 54 macroeconomic indicators. 
Panel A reveals that Stock Market variables, notably new issues and corporate fundamental performance, exhibit the highest predictive power for stock returns. Additionally, variables associated with Consumption (CON), Investment (INV), and Interest and Exchange Rates (IER) demonstrate substantial forecasting capability compared to other groups.
For stock volatility, the lagged stock variance proves to be a robust predictor, aligning with the volatility clustering in financial markets. Beyond the Stock Market conditions, variables pertaining to Interest Rates and Exchange Rates (IER) and Money and Credit (MC) also display considerable predictive relevance for volatility prediction.
Panels C and D indicate that variables related to Prices (PR) and Money and Credit (MC) possess the greatest predictive strength for changes in CPI and PPI, which is anticipated given that several price variables are directly related to indices of consumer and producer prices.

\section{In-Sample Results}\label{sec4}
In this section, we evaluate the in-sample performance of the proposed method by examining the concentration patterns and the loadings of the factors derived from various supervised factor-based techniques. The aim is to clarify the relationship between macroeconomic variables and the variables of interest, providing insights into the predictive capabilities of our method.

We begin by examining the factor concentration patterns across various methods. Specifically, we evaluate the factor extraction results from several approaches: feature-screened and scaled PCA (SSPCA-f, Steps $1+2+3$), dynamic-screened and scaled PCA (SSPCA-d, Steps $1'+2'+3$). These are compared against traditional PCA and two other supervised methods—scaled PCA (sPCA, Steps 2+3) from \cite{scaledpca} and Screened PCA (srPCA, Steps 1+3), which combines correlation-based screening with PCA—due to their relevance to our approach.
\autoref{tab:eigencontribution} presents the eigenvalues for predicting stock returns and CPI, while the results for predicting stock volatility and PPI are shown in \autoref{tab:add_eigen_concentration}. The first factor from PCA explains about 15\% of the total variance, while the first factor from SSPCA accounts for a larger proportion, ranging from 30\% to 63\%. This variation highlights the role of target variables in guiding the PCA process.   Subsequent components, however, contribute less, indicating a diminished explanatory power. Thus, integrating Lasso for factor selection is both logical and crucial for improving forecasting performance. We also observe that SSPCA-f and SSPCA-d lead to different contributions from individual factors, suggesting that the screening and scaling methods extract distinct information from the temporal and feature dimensions, especially when lagged time series are present.

\begin{table}[!ht]
\centering
\renewcommand\arraystretch{1.3}
\caption{Eigenvalues of the  covariance matrices using different PCA techniques.}
\label{tab:eigencontribution}
\resizebox{\textwidth}{!}{
\begin{threeparttable}
    \begin{tabular}{cccccccccccc}
\hline
     &       &  & \multicolumn{4}{c}{Panel A: Excess Stock Return} &  & \multicolumn{4}{c}{Panel C: CPI}  \\ \cline{4-7} \cline{9-12} 
     & PCA   &  & sPCA      & srPCA     & SSPCA-f     & SSPCA-d    &  & sPCA  & srPCA & SSPCA-f & SSPCA-d \\ \hline
1st  & 0.150 &  & 0.611     & 0.226     & 0.631       & 0.603      &  & 0.291 & 0.198 & 0.292   & 0.310   \\
2nd  & 0.102 &  & 0.123     & 0.143     & 0.127       & 0.081      &  & 0.189 & 0.130 & 0.189   & 0.148   \\
3rd  & 0.082 &  & 0.068     & 0.108     & 0.070       & 0.061      &  & 0.119 & 0.095 & 0.119   & 0.113   \\
4th  & 0.074 &  & 0.059     & 0.080     & 0.060       & 0.043      &  & 0.085 & 0.073 & 0.086   & 0.069   \\
5th  & 0.057 &  & 0.033     & 0.076     & 0.034       & 0.033      &  & 0.047 & 0.060 & 0.047   & 0.041   \\
6th  & 0.048 &  & 0.023     & 0.070     & 0.024       & 0.028      &  & 0.045 & 0.046 & 0.045   & 0.038   \\
7th  & 0.040 &  & 0.022     & 0.058     & 0.022       & 0.021      &  & 0.035 & 0.040 & 0.036   & 0.033   \\
8th  & 0.038 &  & 0.015     & 0.054     & 0.013       & 0.020      &  & 0.032 & 0.039 & 0.032   & 0.029   \\
9th  & 0.033 &  & 0.008     & 0.052     & 0.006       & 0.017      &  & 0.026 & 0.035 & 0.026   & 0.025   \\
10th & 0.032 &  & 0.006     & 0.043     & 0.004       & 0.015      &  & 0.020 & 0.031 & 0.020   & 0.022   \\
11th & 0.030 &  & 0.005     & 0.033     & 0.004       & 0.010      &  & 0.018 & 0.027 & 0.018   & 0.020   \\
12th & 0.029 &  & 0.004     & 0.025     & 0.003       & 0.008      &  & 0.014 & 0.024 & 0.014   & 0.015   \\
13th & 0.024 &  & 0.003     & 0.021     & 0.002       & 0.008      &  & 0.013 & 0.023 & 0.013   & 0.015   \\
14th & 0.023 &  & 0.003     & 0.011     & 0.001       & 0.006      &  & 0.010 & 0.021 & 0.010   & 0.013   \\
15th & 0.021 &  & 0.002     & 0.002     & 0.000       & 0.006      &  & 0.009 & 0.019 & 0.009   & 0.012   \\ \hline
\end{tabular}%
\begin{tablenotes}
\item This table presents the eigenvalues of the in-sample covariance matrices in predicting 1-month ahead excess stock return (Panel A) and CPI (Panel C), sorted in descending order and normalized to sum to one. Each eigenvalue indicates the proportion of total variance accounted for by the corresponding principal component.  We compare our factor extraction methods (SSPCA-f, SSPCA-d) with traditional PCA (PCA) and two other supervised techniques: the scaled PCA (sPCA) of \cite{scaledpca} and the screened PCA (srPCA), which combines the screening technique and PCA. We retain the top quartile of predictors to represent the threshold parameter $\kappa_1$ and $\kappa_2$ in the screening step. The sample period is from November 1996 to June 2013.
\end{tablenotes}
\end{threeparttable}%
}
\end{table}

\begin{figure}[htp]
  \includegraphics[width = \textwidth]{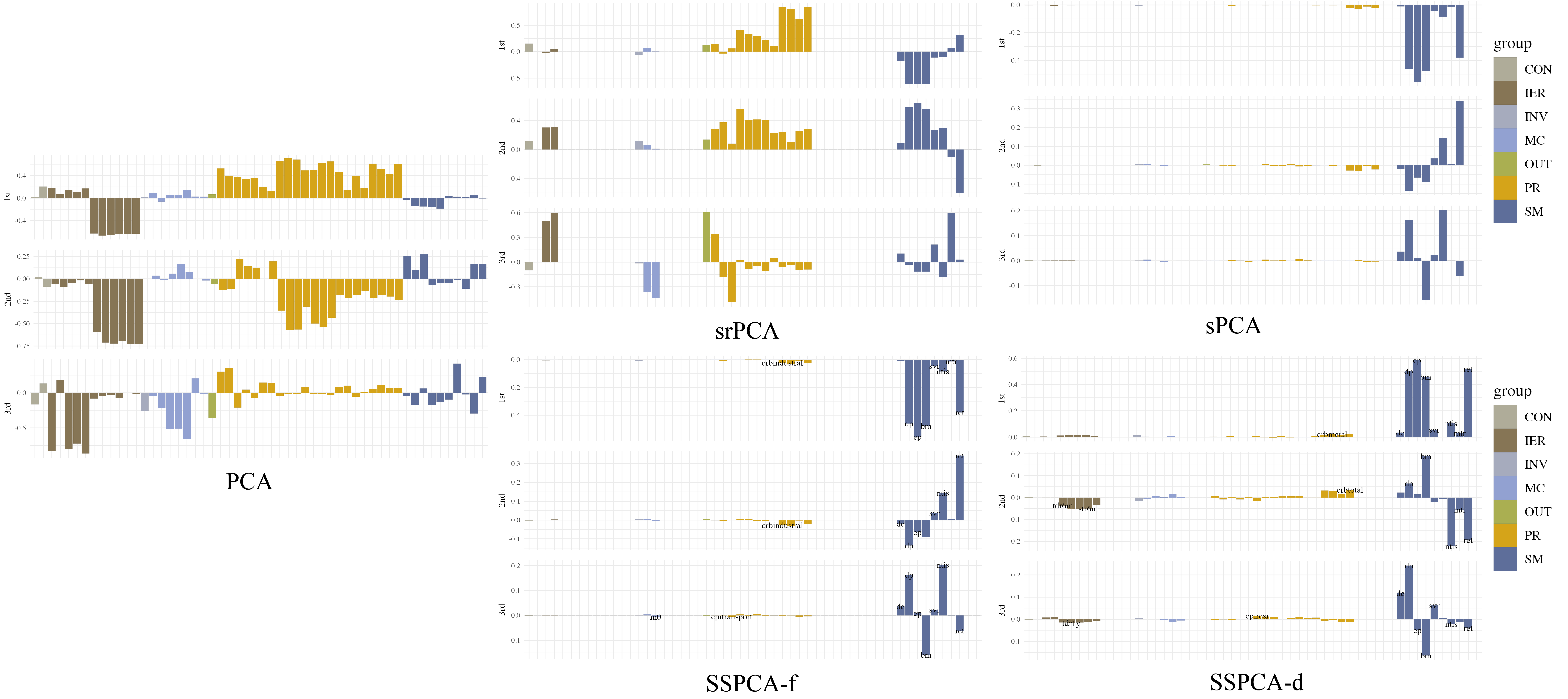}
    \caption{First to third loadings of different PCA factors for predicting 1-month ahead excess stock returns. The five panels represent traditional PCA (PCA), scaled PCA (sPCA) of \cite{scaledpca} and screened PCA (srPCA), feature-screened and scaled PCA (SSPCA-f), dynamic-screened and scaled PCA (SSPCA-d), respectively. The variables are grouped into seven categories: Output (OUT), Stock Market (SM), Prices (PR), Interest Rates and Exchange Rates (IER), Money and Credit (MC), Consumption (CON), and Investment (INV). The sample period is from November 1996 to June 2013.}
    \label{fig:factorcontent}
\end{figure}

Next, we conduct a detailed comparison of the factor composition of the first to third PCA factors guided by factor models just mentioned for predicting 1-month ahead excess stock returns, as shown in \autoref{fig:factorcontent}. Specifically, the initial two factors of traditional PCA capture a diverse array of economic indicators, predominantly correlating with price indices (PR), interest and exchange rates (IER), and also significantly involving money and credit and stock market variables. The third factor is mainly influenced by IER and stock market (SM) indicators.
When using the screening method as a supervisory technique, the factors load primarily on certain price-related (PR) and stock market (SM) variables, followed by INV and IER indicators. In contrast, the scaled PCA allocates most of its factor weights to the stock market, effectively scaling other predictors to negligible values. 
The loading factors derived from the SSPCA-f method closely align with those of scaled PCA. This alignment occurs because the scaled method disproportionately emphasizes critical economic indicators, diminishing the contribution of the screening method in the excess stock return context, especially when few variables are filtered out.  

\begin{figure}[htp]
  \includegraphics[width = \textwidth]{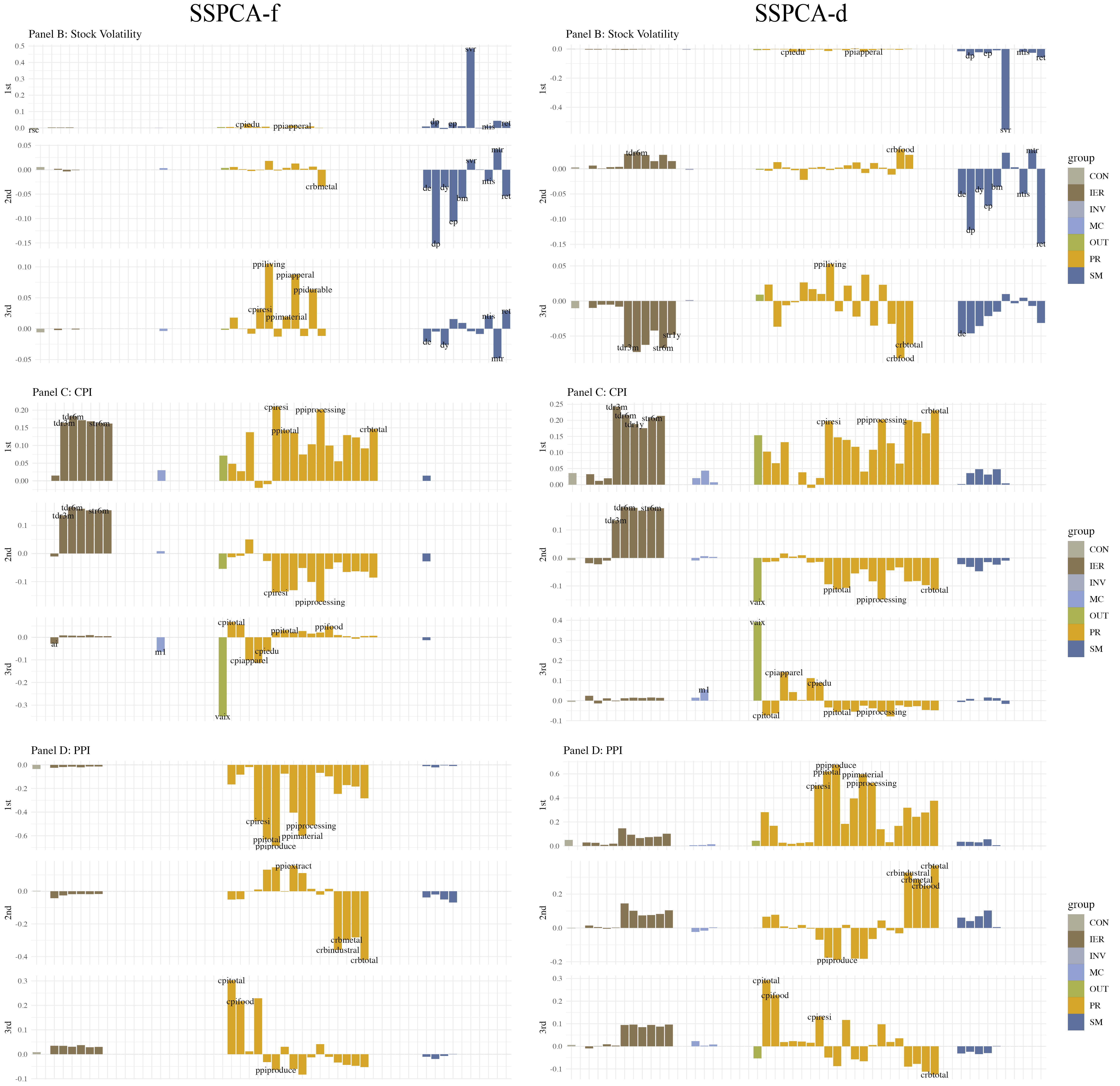}
    \caption{First to third SSRF factor loadings for predicting economic indicators. This figure compares the factor loadings of SSPCA-f and SSPCA-d in predicting 1-month ahead of change in stock volatility (Panel B), change in CPI (Panel C), and change in PPI (Panel D). These variables are grouped into seven categories: Output (OUT), Stock Market (SM), Prices (PR), Interest Rates and Exchange Rates (IER), Money and Credit (MC), Consumption (CON), and Investment (INV).  $\kappa_1$ preserves the top 50\% of predictors ranked by their marginal correlations with the target variable. $\kappa_2$ retains the top 75\% of variables based on their $R^2$ significance. The sample period is 1996:11--2013:06.}
    \label{fig:factorcontent2}
\end{figure}

While SSPCA-d and SSPCA-f differ in coefficient signs, both methods amplify  similar core information stock market (SM) in the primary components.  At the same time, their selection on less dominant factors varies, offering complementary insights, which can enhance predictive performance.
\autoref{fig:factorcontent2} presents additional examples of factor loadings from SSPCA-f and SSPCA-d for forecasting 1-month ahead changes in stock volatility, CPI, and PPI, revealing distinct patterns compared to traditional PCA (as shown in \autoref{fig:factorcontent}). 
In Panel B, the loadings emphasize SM variables such as stock variance and the inflation rate, underscoring their relevance in forecasting stock volatility.
For CPI prediction, the first factor is primarily associated with PR and IER. While these variables align with those identified by PCA, their weights are adjusted, filtering out many less significant ones. Notably, the output (OUT) group is strongly amplified in both SSPCA-f and SSPCA-d. For PPI, the loadings are closely linked to specific price variables, which is expected given their inherent direct relationship. 

\section{Out-of-Sample Results}\label{sec5}
In this section, we evaluate the out-of-sample predictive performance of SSRF in the macroeconomic prediction task using MSFE in (\ref{msfe}), with the evaluation period ranging from $t=$ June 2013 to $T=$ December 2019. Given that our methodology is grounded in the principles of sPCA, correlation learning, and regression, we evaluate and compare the individual and combined performances of these methods with the proposed SSRF approach. 

We first compare SSRF with various dimension reduction techniques: PCA, sPCA, srPCA, and SSPCA-f in conjunction with both Linear Regression (LM) and Lasso regression techniques in forecasting. \autoref{tab:ooscompare} presents the out-of-sample MSFE for predicting various economic indicators using these methods, with a particular focus on the three factor space conditions introduced in Section \ref{sec2}. Specifically, SSRF(1) to SSRF(3) correspond to applying Lasso regression to: (1) $\bF_{f}$, (2) $\bF_{d}$ and (3) $\bF_{hyb}$. 
We treat (1) and (2) as baseline methods, since they rely on single-dimensional scaling techniques and commonly used screening and regularization frameworks. Our focus is on evaluating (3), which examines how the SSRF model integrates information from both feature and temporal dimensions to enhance predictive performance.

The results in \autoref{tab:ooscompare} show that SSRF consistently achieves a comparatively low MSFE, demonstrating superior predictive performance across most economic indicators and forecast horizons. Notably, SSRF(3) outperforms alternative methods in all scenarios except for stock volatility prediction. This highlights the advantage of integrating signals from both dimensions, which capture complementary predictive information, as supported by the factor loading results in the previous section. For stock volatility, however, the SSRF approach does not surpass the simpler srPCA+Lasso method, which excludes the scaling step.
We attribute this divergence to two key factors: (1) the screening step in SSRF already isolates a dominant predictor that largely governs volatility dynamics, and (2) the subsequent scaling procedure in sPCA may inadvertently amplify noise rather than enhancing signal discrimination. 
Furthermore, as the forecast horizon $h$ extends, the prediction accuracy gradually declines. This is understandable, as uncertainty increases with a longer forecasting time horizon, making accurate predictions more challenging.

\begin{table}[htb]\scriptsize
\renewcommand\arraystretch{1.3}
\caption{Out-of-sample MSFE produced by different methods.\\}\label{tab:ooscompare}
\centering\resizebox{\textwidth}{!}{
\begin{threeparttable}
\begin{tabular}{cccccccccc}
\hline
                       &             &  & $h=1$          & $h=2$          & $h=3$          &  & $h=1$          & $h=2$          & $h=3$          \\ \cline{3-6} \cline{8-10}
                       &             & \multicolumn{4}{c}{Panel A: Excess Stock Return}    & \multicolumn{4}{c}{Panel B: Stock Volatility}       \\ \hline
\multirow{4}{*}{LM}    & PCA+LM      &  & 0.764          & 0.774          & 0.787          &  & 0.795          & 0.833          & 0.886          \\
                       & sPCA+LM     &  & 0.642          & 0.823          & 0.806          &  & 0.818          & 0.925          & 1.062          \\
                       & srPCA+LM    &  & 0.718          & 0.827          & 0.840          &  & 0.788          & 0.935          & 1.156          \\
                       & SSPCA-f+LM  &  & 0.640          & 0.809          & 0.789          &  & 0.805          & 0.895          & 1.032          \\ \cline{1-2} \cline{4-6} \cline{8-10} 
\multirow{6}{*}{LASSO} & PCA+LASSO   &  & 0.684          & 0.781          & 0.798          &  & 0.775          & 0.832          & 0.888          \\
                       & sPCA+LASSO  &  & 0.638          & 0.798          & 0.796          &  & 0.806          & 0.834          & 1.023          \\
                       & srPCA+LASSO &  & 0.637          & 0.775          & 0.790          &  & \textbf{0.772} & \textbf{0.803} & 0.885          \\ \cline{2-2} \cline{4-6} \cline{8-10} 
                       & SSRF(1)     &  & 0.633          & 0.767          & 0.782          &  & 0.799          & 0.829          & 0.905          \\
                       & SSRF(2)     &  & 0.639          & 0.784          & 0.783          &  & 0.842          & 0.834          & \textbf{0.868} \\
                       & SSRF(3)     &  & \textbf{0.612} & \textbf{0.760} & \textbf{0.758} &  & 0.782          & 0.859          & 0.900          \\ \hline
                       &             &  & $h=1$          & $h=2$          & $h=3$          &  & $h=1$          & $h=2$          & $h=3$          \\ \cline{3-6} \cline{8-10} 
                       &             & \multicolumn{4}{c}{Panel C: CPI}                    & \multicolumn{4}{c}{Panel D: PPI}                    \\ \hline
\multirow{4}{*}{LM}    & PCA+LM      &  & 0.762          & 0.760          & 0.763          &  & 0.566          & 0.753          & 0.819          \\
                       & sPCA+LM     &  & 0.755          & 0.779          & 0.820          &  & 0.533          & 0.753          & 0.813          \\
                       & srPCA+LM    &  & 0.707          & 0.835          & 0.807          &  & 0.570          & 0.751          & 0.821          \\
                       & SSPCA-f+LM  &  & 0.701          & 0.767          & 0.795          &  & 0.552          & 0.752          & 0.803          \\ \cline{1-2} \cline{4-6} \cline{8-10} 
\multirow{6}{*}{LASSO} & PCA+LASSO   &  & 0.787          & 0.748          & 0.776          &  & 0.578          & 0.755          & 0.804          \\
                       & sPCA+LASSO  &  & 0.729          & 0.793          & 0.802          &  & 0.532          & 0.741          & 0.796          \\
                       & srPCA+LASSO &  & 0.707          & 0.740          & 0.774          &  & 0.526          & 0.726          & 0.789          \\ \cline{2-2} \cline{4-6} \cline{8-10} 
                       & SSRF(1)     &  & 0.698          & 0.732          & \textbf{0.758} &  & 0.524          & 0.728          & 0.794          \\
                       & SSRF(2)     &  & 0.691          & 0.759          & 0.774          &  & 0.527          & 0.722          & 0.793          \\
                       & SSRF(3)     &  & \textbf{0.687} & \textbf{0.726} & 0.772          &  & \textbf{0.514} & \textbf{0.721} & \textbf{0.781} \\ \hline
\end{tabular}%
\begin{tablenotes}
 \item This table presents the out-of-sample MSFE for predicting excess stock return (Panel A), change in stock volatility (Panel B), change in CPI (Panel C), and change in PPI (Panel D), using 54 macroeconomic variables over forecast horizons of 1 to 3 months ($h=1,2,3$). We compare our proposed method (SSRF) with various combinations of regression techniques (Linear Regression (LM) versus Lasso) and dimension reduction methods: traditional PCA (PCA), scaled PCA (sPCA), the screened PCA (srPCA) as well as their combination (SSPCA-f). SSRF(1)-(3) correspond to the Lasso regression applied to the three factor spaces outlined in Section \ref{sec2}. For the Linear Regression model (LM), the ratio-based estimation method is used to determine the number of factors. The out-of-sample testing period extends is 2013:07-2019:12.
\end{tablenotes}
\end{threeparttable}%
}
\end{table}

The ratio-based approach typically reduces PCA factor selection to a lower dimension. In contrast, SSRF opts for a more inclusive initial selection of factors, denoted by a potentially larger number $r$. This strategy proves particularly beneficial in the subsequent phase of the SSRF procedure, where Lasso regression is employed. A comparative analysis between the Linear Regression (LM) and Lasso regression, as shown in  \autoref{tab:ooscompare}, clearly demonstrates the proficiency of Lasso in reducing forecast errors. Further investigation examines how the out-of-sample performance varies with the numbers of factors. As shown in \autoref{tab:diffr}, the SSRF method consistently performs well and remains stable regardless of the number of latent factors.
This finding suggests that when there is uncertainty regarding the optimal number of factor dimensions, a prudent strategy is to select a more expansive factor space and employ penalized regression methods like Lasso for automatic variable selection. 

\begin{figure}[htb]
  \includegraphics[width = \textwidth]{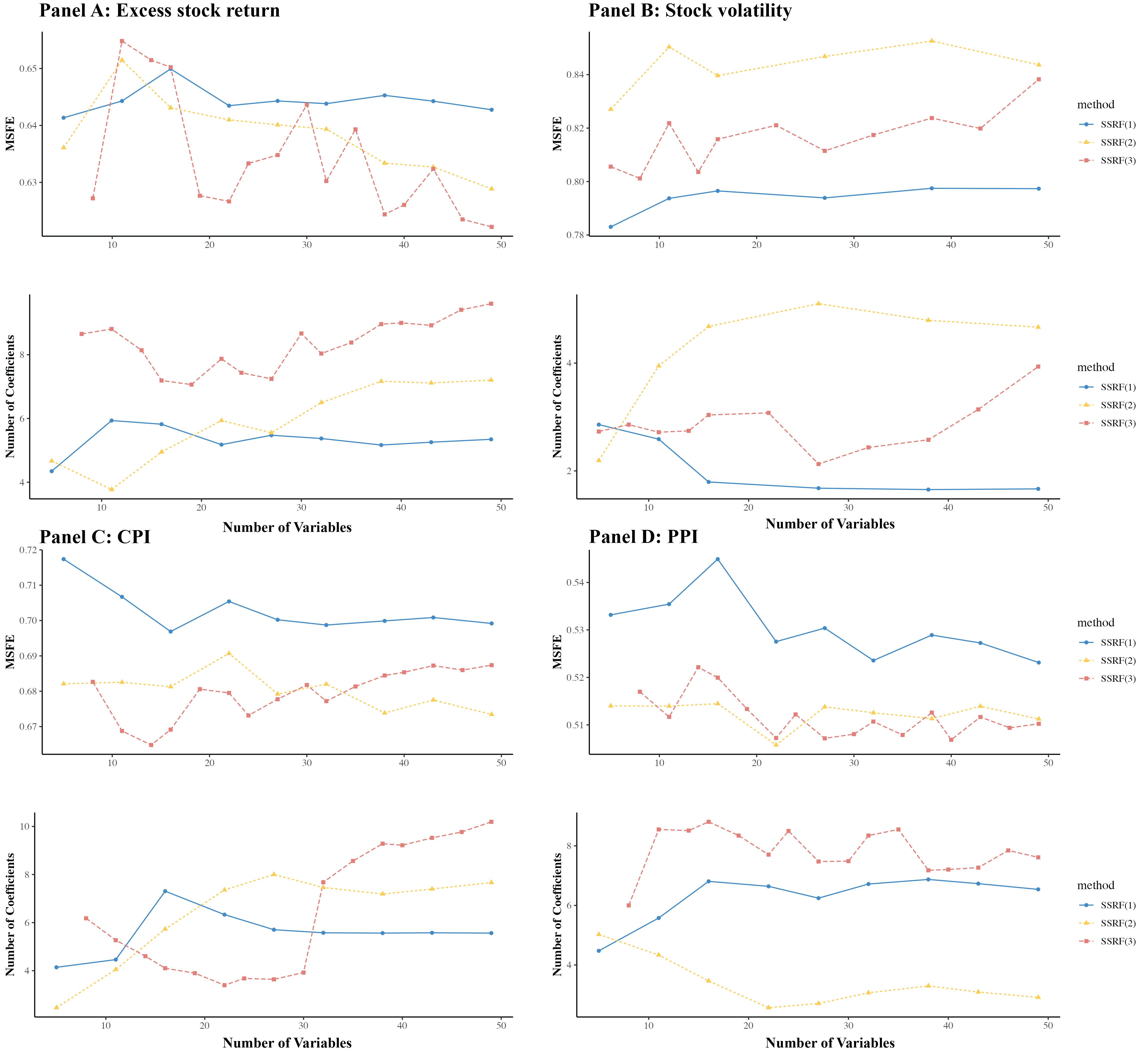}
    \caption{The impact of screening parameters on MSFE and the number of factors selected by Lasso. The horizontal axis represents the number of retained variables.  The four panels represent 1-month ahead forecasting of excess stock return (Panel A), stock volatility (Panel B),  CPI (Panel C), and PPI (Panel D). SSRF(1)-(3) correspond to Lasso regressions applied to the three factor spaces described in Section \ref{sec2}. The number of PCA factors is fixed at 10.}
    \label{fig:kappachange}
\end{figure}

We further investigate how the screening parameter influences both MSFE and the number of factors selected by the Lasso models, as shown in \autoref{fig:kappachange}. Specifically, $\kappa_1$ is derived from feature relevance, while $\kappa_2$ is based on the dynamic regression fit. As $\kappa_1,\kappa_2$ decreases, leading to an increase in the number of original variables, the number of factors selected by Lasso becomes stable except for SSRF(3), as many variables are irrelevant and do not contribute to predictive power.
The SSRF(3) model generally selects more factors than other configurations. However, this increase is not a simple doubling, as some selected factors may contain redundant information.

For the prediction of excess stock returns in Panel A, SSRF(3) achieves the lowest MSFE, indicating that integrating cross-sectional and temporal dimensions enhances signal extraction efficiency. This performance improvement occurs when retaining a higher proportion of factors, suggesting that forecasting stock returns using macroeconomic data operates in a low signal-to-noise environment. In such settings, a broader set of variables is necessary to capture potential predictive information effectively. In contrast, stock volatility forecasting reveals a different outcome, where SSRF(1) delivers superior accuracy with minimal retained variables, reflecting the dominance of a single strong predictor in volatility dynamics. This result aligns with established empirical evidence: while volatility exhibits time-series clustering patterns, macroeconomic factors exert limited explanatory power at low frequencies.
For CPI and PPI forecasting, SSRF(2) outperforms SSRF(1), emphasizing the benefits of incorporating temporal information from dynamic factors. This result highlights the critical role of temporal persistence in capturing price index dynamics and enhancing forecast accuracy.

\begin{table}[htb]
\renewcommand\arraystretch{1.5}
\caption{Out-of-sample MSFE with varying number of factors.\\}\label{tab:diffr}
\centering\resizebox{1\textwidth}{!}{
\begin{threeparttable}
\begin{tabular}{cccccccccccccccc}
\hline
\multicolumn{1}{l}{} & \multicolumn{7}{c}{$h=1$}                                                                                                                                                                              &                      & \multicolumn{7}{c}{$h=3$}                                                                                                                                                                              \\ \cline{2-8} \cline{10-16} 
\multicolumn{1}{l}{} & \multicolumn{1}{l}{$r=5$} & \multicolumn{1}{l}{$r=7$} & \multicolumn{1}{l}{$r=10$} & \multicolumn{1}{l}{$r=15$} & \multicolumn{1}{l}{$r=20$} & \multicolumn{1}{l}{$r=25$} & \multicolumn{1}{l}{$r=30$} & \multicolumn{1}{l}{} & \multicolumn{1}{l}{$r=5$} & \multicolumn{1}{l}{$r=7$} & \multicolumn{1}{l}{$r=10$} & \multicolumn{1}{l}{$r=15$} & \multicolumn{1}{l}{$r=20$} & \multicolumn{1}{l}{$r=25$} & \multicolumn{1}{l}{$r=30$} \\ \hline
\multicolumn{16}{l}{Panel A: Prediction of Excess Stock Return}                                                                                                                                                                                                                                                                                                                                                                                               \\
PCA                  & 0.648                     & 0.654                     & 0.689                      & 0.694                      & 0.678                      & 0.666                      & 0.684                      &                      & 0.785                     & 0.783                     & 0.783                      & 0.798                      & 0.795                      & 0.786                      & 0.785                      \\
sPCA                 & 0.643                     & 0.641                     & 0.654                      & 0.659                      & 0.647                      & 0.648                      & 0.638                      &                      & 0.787                     & 0.799                     & 0.802                      & 0.796                      & 0.801                      & 0.799                      & 0.804                      \\
SrPCA                & 0.639                     & 0.640                     & 0.637                      & 0.646                      & 0.647                      & 0.644                      & 0.637                      &                      & 0.791                     & 0.779                     & 0.787                      & 0.790                      & 0.783                      & 0.784                      & 0.779                      \\
SSRF(1)              & 0.639                     & 0.637                     & 0.638                      & 0.647                      & 0.638                      & 0.636                      & 0.633                      &                      & 0.780                     & 0.774                     & 0.781                      & 0.782                      & 0.778                      & 0.775                      & 0.777                      \\
SSRF(2)              & 0.650                     & 0.638                     & 0.627                      & 0.648                      & 0.646                      & 0.642                      & 0.639                      &                      & 0.793                     & 0.776                     & 0.795                      & 0.783                      & 0.775                      & 0.775                      & 0.772                      \\
SSRF(3)              & \textbf{0.626}            & \textbf{0.636}            & \textbf{0.625}             & \textbf{0.646}             & \textbf{0.639}             & \textbf{0.625}             & \textbf{0.612}             &                      & \textbf{0.786}            & \textbf{0.770}            & \textbf{0.763}             & \textbf{0.758}             & \textbf{0.762}             & \textbf{0.767}             & \textbf{0.769}             \\ \hline
\multicolumn{16}{l}{Panel B: Prediction of Stock Volatility}                                                                                                                                                                                                                                                                                                                                                                                                  \\
PCA                  & 0.775                     & 0.777                     & \textbf{0.775}             & 0.800                      & 0.795                      & 0.806                      & \textbf{0.782}             &                      & 0.886                     & 0.887                     & 0.887                      & 0.886                      & \textbf{0.885}             & 0.888                      & 0.887                      \\
sPCA                 & 0.806                     & 0.808                     & 0.809                      & 0.804                      & 0.811                      & 0.808                      & 0.807                      &                      & 1.052                     & 1.032                     & 1.062                      & 1.070                      & 1.064                      & 1.023                      & 1.015                      \\
SrPCA                & \textbf{0.772}            & \textbf{0.762}            & 0.787                      & \textbf{0.772}             & \textbf{0.770}             & \textbf{0.787}             & 0.783                      &                      & \textbf{0.885}            & \textbf{0.886}            & \textbf{0.886}             & \textbf{0.874}             & \textbf{0.885}             & 0.885                      & \textbf{0.884}             \\
SSRF(1)              & 0.799                     & 0.799                     & 0.794                      & 0.790                      & 0.805                      & 0.796                      & 0.789                      &                      & 0.996                     & 0.975                     & 0.960                      & 0.993                      & 0.916                      & 0.905                      & 0.912                      \\
SSRF(2)              & 0.842                     & 0.847                     & 0.847                      & 0.829                      & 0.838                      & 0.814                      & 0.805                      &                      & 1.041                     & 0.987                     & 0.934                      & 0.900                      & 0.912                      & \textbf{0.868}             & 0.895                      \\
SSRF(3)              & 0.782                     & 0.787                     & 0.787                      & 0.802                      & 0.809                      & 0.819                      & 0.812                      &                      & 0.981                     & 0.967                     & 0.934                      & 0.907                      & 0.901                      & 0.900                      & 0.905                      \\ \hline
\multicolumn{16}{l}{Panel C: Prediction of CPI}                                                                                                                                                                                                                                                                                                                                                                                                               \\
PCA                  & 0.724                     & 0.722                     & 0.720                      & 0.720                      & 0.727                      & 0.787                      & 0.749                      &                      & 0.795                     & 0.794                     & 0.789                      & 0.781                      & 0.776                      & 0.787                      & 0.787                      \\
sPCA                 & 0.716                     & 0.698                     & 0.703                      & 0.705                      & 0.709                      & 0.729                      & 0.724                      &                      & 0.819                     & 0.799                     & 0.805                      & 0.792                      & 0.802                      & 0.808                      & 0.803                      \\
SrPCA                & 0.706                     & 0.695                     & 0.697                      & 0.701                      & 0.712                      & 0.707                      & 0.708                      &                      & 0.770                     & 0.773                     & 0.775                      & 0.767                      & 0.774                      & 0.768                      & 0.773                      \\
SSRF(1)              & 0.710                     & 0.692                     & 0.697                      & 0.694                      & 0.698                      & 0.695                      & 0.692                      &                      & 0.766                     & \textbf{0.767}            & \textbf{0.767}             & \textbf{0.758}             & \textbf{0.758}             & \textbf{0.765}             & \textbf{0.770}             \\
SSRF(2)              & 0.689                     & 0.679                     & 0.676                      & 0.689                      & 0.691                      & 0.725                      & 0.720                      &                      & 0.783                     & 0.778                     & 0.776                      & 0.773                      & 0.774                      & 0.772                      & 0.777                      \\
SSRF(3)              & \textbf{0.676}            & \textbf{0.678}            & \textbf{0.677}             & \textbf{0.678}             & \textbf{0.687}             & \textbf{0.694}             & \textbf{0.690}             &                      & \textbf{0.764}            & 0.772                     & 0.772                      & 0.763                      & 0.772                      & 0.770                      & 0.779                      \\ \hline
\multicolumn{16}{l}{Panel D:Prediction of PPI}                                                                                                                                                                                                                                                                                                                                                                                                                \\
PCA                  & 0.578                     & 0.591                     & 0.554                      & 0.532                      & 0.532                      & 0.539                      & 0.530                      &                      & 0.818                     & 0.804                     & 0.799                      & 0.805                      & 0.825                      & 0.816                      & 0.816                      \\
sPCA                 & 0.532                     & 0.531                     & 0.529                      & 0.558                      & 0.562                      & 0.553                      & 0.551                      &                      & 0.791                     & 0.796                     & 0.807                      & 0.812                      & 0.804                      & 0.800                      & 0.787                      \\
SrPCA                & 0.526                     & 0.532                     & 0.535                      & 0.528                      & 0.532                      & 0.530                      & 0.522                      &                      & 0.794                     & 0.789                     & 0.794                      & 0.792                      & 0.792                      & 0.786                      & 0.784                      \\
SSRF(1)              & 0.524                     & 0.525                     & 0.526                      & 0.537                      & 0.533                      & 0.530                      & 0.531                      &                      & 0.789                     & 0.794                     & 0.792                      & 0.790                      & \textbf{0.788}             & \textbf{0.784}             & \textbf{0.787}             \\
SSRF(2)              & 0.527                     & \textbf{0.510}            & \textbf{0.511}             & \textbf{0.512}             & \textbf{0.512}             & \textbf{0.511}             & \textbf{0.512}             &                      & 0.794                     & 0.793                     & 0.792                      & 0.795                      & 0.796                      & 0.798                      & 0.800                      \\
SSRF(3)              & \textbf{0.514}            & 0.519                     & 0.526                      & 0.525                      & 0.526                      & 0.526                      & 0.536                      &                      & \textbf{0.783}            & \textbf{0.781}            & \textbf{0.786}             & \textbf{0.788}             & 0.789                      & 0.794                      & 0.792                      \\ \cline{1-16}
\end{tabular}%
\begin{tablenotes}
\item This table presents the out-of-sample root mean squared forecast errors (MSFE) for predicting economic indicators as the number of factors changes. Specifically, we compare the performance of SSRF with other factor-based methods—PCA, sPCA and srPCA—combined with Lasso in predicting 1-month ahead ($h=1$) and 3-month ahead ($h=3$) excess stock return (Panel A), change in stock volatility (Panel B), change in CPI (Panel C), and change in PPI (Panel D) using 54 macroeconomic variables. 
\end{tablenotes}
\end{threeparttable}
}
\end{table} 

We expand our comparison to include other established machine learning techniques suitable for handling many predictors.  Specifically, we consider Partial Least Squares (PLS) as introduced by  \cite{wold1966estimation}, the Smoothly Clipped Absolute Deviation (SCAD) method proposed by  \cite{fan2001variable}, and the Elastic Net (ENet) regularization technique developed by \cite{zou2005regularization}, respectively.  The empirical results in \autoref{tab:ooscompare2} reveal that the SSRF method generally exhibits the lowest MSFE values across most target variables and horizons, suggesting its superior forecasting capabilities. Competing methods, such as SCAD and Elastic Net, also demonstrate strong forecasting performance.

 \begin{table}[htb]
\centering
\renewcommand\arraystretch{1.3}
\caption{Comparison of out-of-sample MSFE by different methods.\\}\label{tab:ooscompare2}
\resizebox{0.8\textwidth}{!}{%
\begin{threeparttable}
\begin{tabular}{ccccccccc}
\hline
        &  & $h=1$          & $h=2$          & $h=3$          &  & $h=1$          & $h=2$          & $h=3$          \\ \hline
        & \multicolumn{4}{c}{Panel A: Excess Stock Return}    & \multicolumn{4}{c}{Panel B: Stock Volatility}       \\ \hline
PLS     &  & 0.645          & 0.815          & 0.803          &  & 0.788          & 0.873          & 0.985          \\
SCAD    &  & 0.648          & 0.814          & 0.812          &  & 0.802          & 0.881          & 1.098          \\
ENet    &  & 0.648          & 0.775          & 0.783          &  & 0.803          & 0.861          & 0.970          \\
LASSO   &  & 0.643          & 0.782          & 0.785          &  & 0.799          & 0.832          & 0.946          \\
SSRF(1) &  & 0.633          & 0.767          & 0.782          &  & 0.799          & \textbf{0.829} & 0.905          \\
SSRF(2) &  & 0.639          & 0.784          & 0.783          &  & 0.842          & 0.834          & \textbf{0.868} \\
SSRF(3) &  & \textbf{0.612} & \textbf{0.760} & \textbf{0.758} &  & \textbf{0.782} & 0.859          & 0.900          \\ \hline
        &  & $h=1$          & $h=2$          & $h=3$          &  & $h=1$          & $h=2$          & $h=3$          \\ \hline
        & \multicolumn{4}{c}{Panel C: CPI}                    & \multicolumn{4}{c}{Panel D: PPI}                    \\ \hline
PLS     &  & 0.757          & 0.772          & 0.837          &  & 0.547          & 0.753          & 0.812          \\
SCAD    &  & 0.691          & 0.752          & 0.766          &  & 0.537          & 0.792          & 0.795          \\
ENet    &  & 0.719          & 0.747          & 0.770          &  & 0.536          & 0.733          & 0.799          \\
LASSO   &  & 0.735          & 0.765          & 0.777          &  & 0.544          & 0.738          & 0.797          \\
SSRF(1) &  & 0.698          & 0.732          & \textbf{0.758} &  & 0.524          & 0.728          & 0.794          \\
SSRF(2) &  & 0.691          & 0.759          & 0.774          &  & 0.527          & 0.722          & 0.793          \\
SSRF(3) &  & \textbf{0.687} & \textbf{0.726} & 0.772          &  & \textbf{0.514} & \textbf{0.721} & \textbf{0.781} \\ \hline
\end{tabular}%
\begin{tablenotes}
\item The table provides a detailed analysis of the out-of-sample mean square prediction error (MSFE) for predicting key economic indicators such as excess stock return, stock volatility, CPI (Consumer Price Index), and PPI (Producer Price Index) using macroeconomic variables using 54 macro-economic variables. We compared with several methods, including PLS, SCAD, and Elastic-net (ENet).
In the implementation of PLS, we utilize a ratio-based method to select the factor number. For Elastic-net (ENet), we adopt a rigorous threefold cross-validation approach to estimate forecasts. 
\end{tablenotes}
\end{threeparttable}
}
\end{table}

Overall, the empirical analysis of China's macroeconomic indices suggests that the proposed SSRF method  outperforms many commonly used factor-based and linear approaches in most cases. Furthermore, the inclusion of regularization enhances out-of-sample forecasting performance compared to traditional methods that do not incorporate regularization.

\section{Conclusion}\label{sec6}

In this paper, we introduced a Supervised Screening and Regularized Factor-based (SSRF) forecasting method designed for economic forecasting in high-dimensional data contexts. Compared to existing methods, SSRF maximizes the utilization of information between target variables and predictors in both feature and temporal dimensions, providing an effective approach for economic forecasting using supervised machine learning techniques. Our method integrates several useful modeling techniques, including correlation learning, scaled PCA, factor modeling, and forecasting. This structured approach is computationally efficient and easy to implement.

Empirically, we apply the proposed SSRF method to predict some key Chinese macroeconomic indices using a dataset of monthly macroeconomic variables from China. Our method demonstrates predictive advantages over several commonly used benchmark methods in the literature. The proposed method offers another valuable option in the toolkit for practitioners interested in out-of-sample economic forecasting with numerous predictors.

\bibliographystyle{ecta}
\addcontentsline{toc}{section}{\refname}
\bibliography{SSRF}

\begin{thebibliography}{22}
\newcommand{\enquote}[1]{``#1''}
\expandafter\ifx\csname natexlab\endcsname\relax\def\natexlab#1{#1}\fi

\bibitem[\protect\citeauthoryear{Bai and Ng}{Bai and Ng}{2002}]{bai2002determining}
\textsc{Bai, J. and S.~Ng} (2002): \enquote{Determining the number of factors in approximate factor models,} \emph{Econometrica}, 70, 191--221.

\bibitem[\protect\citeauthoryear{Bai and Ng}{Bai and Ng}{2008}]{bai2008forecasting}
---\hspace{-.1pt}---\hspace{-.1pt}--- (2008): \enquote{Forecasting economic time series using targeted predictors,} \emph{Journal of Econometrics}, 146, 304--317.

\bibitem[\protect\citeauthoryear{Bai and Ng}{Bai and Ng}{2023}]{bai2023approximate}
---\hspace{-.1pt}---\hspace{-.1pt}--- (2023): \enquote{Approximate factor models with weaker loadings,} \emph{Journal of Econometrics}, 235, 1893--1916.

\bibitem[\protect\citeauthoryear{Bickel and Levina}{Bickel and Levina}{2008}]{bickel2008regularized}
\textsc{Bickel, P.~J. and E.~Levina} (2008): \enquote{Regularized estimation of large covariance matrices,} \emph{The Annals of Statistics}, 36, 199--227.

\bibitem[\protect\citeauthoryear{Fan and Li}{Fan and Li}{2001}]{fan2001variable}
\textsc{Fan, J. and R.~Li} (2001): \enquote{Variable selection via nonconcave penalized likelihood and its oracle properties,} \emph{Journal of the American statistical Association}, 96, 1348--1360.

\bibitem[\protect\citeauthoryear{Fan, Liao, and Mincheva}{Fan et~al.}{2013}]{fan2013thresh}
\textsc{Fan, J., Y.~Liao, and M.~Mincheva} (2013): \enquote{{Large Covariance Estimation by Thresholding Principal Orthogonal Complements},} \emph{Journal of the Royal Statistical Society Series B: Statistical Methodology}, 75, 603--680.

\bibitem[\protect\citeauthoryear{Fan and Lv}{Fan and Lv}{2008}]{fan2008sure}
\textsc{Fan, J. and J.~Lv} (2008): \enquote{Sure independence screening for ultrahigh dimensional feature space,} \emph{Journal of the Royal Statistical Society Series B: Statistical Methodology}, 70, 849--911.

\bibitem[\protect\citeauthoryear{Frank and Friedman}{Frank and Friedman}{1993}]{frank1993statistical}
\textsc{Frank, L.~E. and J.~H. Friedman} (1993): \enquote{A statistical view of some chemometrics regression tools,} \emph{Technometrics}, 35, 109--135.

\bibitem[\protect\citeauthoryear{Gao and Tsay}{Gao and Tsay}{2024}]{gao2024supervised}
\textsc{Gao, Z. and R.~S. Tsay} (2024): \enquote{Supervised dynamic pca: Linear dynamic forecasting with many predictors,} \emph{Journal of the American Statistical Association}, Forthcoming.

\bibitem[\protect\citeauthoryear{Goulet~Coulombe, Leroux, Stevanovic, and Surprenant}{Goulet~Coulombe et~al.}{2022}]{goulet2022machine}
\textsc{Goulet~Coulombe, P., M.~Leroux, D.~Stevanovic, and S.~Surprenant} (2022): \enquote{How is machine learning useful for macroeconomic forecasting?} \emph{Journal of Applied Econometrics}, 37, 920--964.

\bibitem[\protect\citeauthoryear{Huang, Jiang, Li, Tong, and Zhou}{Huang et~al.}{2022}]{scaledpca}
\textsc{Huang, D., F.~Jiang, K.~Li, G.~Tong, and G.~Zhou} (2022): \enquote{Scaled PCA: A New Approach to Dimension Reduction,} \emph{Management Science}, 68, 1678--1695.

\bibitem[\protect\citeauthoryear{Huang and Tsay}{Huang and Tsay}{2024}]{huang2024time}
\textsc{Huang, S.-C. and R.~S. Tsay} (2024): \enquote{Time Series Forecasting with Many Predictors,} \emph{Mathematics}, 12, 2336.

\bibitem[\protect\citeauthoryear{Lam and Yao}{Lam and Yao}{2012}]{lam2012factor}
\textsc{Lam, C. and Q.~Yao} (2012): \enquote{Factor modeling for high-dimensional time series: inference for the number of factors,} \emph{The Annals of Statistics}, 694--726.

\bibitem[\protect\citeauthoryear{Ledoit and Wolf}{Ledoit and Wolf}{2004}]{ledoit2004well}
\textsc{Ledoit, O. and M.~Wolf} (2004): \enquote{A well-conditioned estimator for large-dimensional covariance matrices,} \emph{Journal of multivariate analysis}, 88, 365--411.

\bibitem[\protect\citeauthoryear{Ling, Ma, Tu, and Xie}{Ling et~al.}{2023}]{tu2023}
\textsc{Ling, B., C.~Ma, Y.~Tu, and X.~Xie} (2023): \enquote{Mutation analysis of China 's macroeconomic structure based on high-dimensional factor model (in Chinese),} \emph{Quarterly Journal of Economics and Management}, 2, 37--62.

\bibitem[\protect\citeauthoryear{McCracken and Ng}{McCracken and Ng}{2016}]{mccracken2016fred}
\textsc{McCracken, M.~W. and S.~Ng} (2016): \enquote{FRED-MD: A monthly database for macroeconomic research,} \emph{Journal of Business \& Economic Statistics}, 34, 574--589.

\bibitem[\protect\citeauthoryear{Onatski}{Onatski}{2012}]{onatski2012asymptotics}
\textsc{Onatski, A.} (2012): \enquote{Asymptotics of the principal components estimator of large factor models with weakly influential factors,} \emph{Journal of Econometrics}, 168, 244--258.

\bibitem[\protect\citeauthoryear{Stock and Watson}{Stock and Watson}{2002{\natexlab{a}}}]{stock2002forecasting}
\textsc{Stock, J.~H. and M.~W. Watson} (2002{\natexlab{a}}): \enquote{Forecasting using principal components from a large number of predictors,} \emph{Journal of the American statistical association}, 97, 1167--1179.

\bibitem[\protect\citeauthoryear{Stock and Watson}{Stock and Watson}{2002{\natexlab{b}}}]{stock2002macroeconomic}
---\hspace{-.1pt}---\hspace{-.1pt}--- (2002{\natexlab{b}}): \enquote{Macroeconomic forecasting using diffusion indexes,} \emph{Journal of Business \& Economic Statistics}, 20, 147--162.

\bibitem[\protect\citeauthoryear{Tibshirani}{Tibshirani}{1996}]{tibshirani1996regression}
\textsc{Tibshirani, R.} (1996): \enquote{Regression shrinkage and selection via the lasso,} \emph{Journal of the Royal Statistical Society Series B: Statistical Methodology}, 58, 267--288.

\bibitem[\protect\citeauthoryear{Wold}{Wold}{1966}]{wold1966estimation}
\textsc{Wold, H.} (1966): \enquote{Estimation of principal components and related models by iterative least squares,} \emph{Multivariate analysis}, 391--420.

\bibitem[\protect\citeauthoryear{Zou and Hastie}{Zou and Hastie}{2005}]{zou2005regularization}
\textsc{Zou, H. and T.~Hastie} (2005): \enquote{Regularization and variable selection via the elastic net,} \emph{Journal of the Royal Statistical Society Series B: Statistical Methodology}, 67, 301--320.

\end{thebibliography}
 
\clearpage
\setcounter{page}{1}
\appendix
\begin{center}
    {\bf \Large Online Appendix}
\end{center}

\section{Data Description}
This appendix first lists the 54 macroeconomic time series obtained from
the National Bureau of Statistics, the Wind Economic Database and CSMAR. The last column specifies the group of each series: Output (OUT), Stock Market (SM), Prices (PR), Interest Rates and Exchange Rates (IER), Money and Credit (MC), Consumption (CON), and Investment (INV). The column tcode denotes the following data transformation for a series $x$: (1) no transformation; (2)$\Delta x_t$; (3) $\Delta^2 x_t$; (4) $\ln(x_t)$; (5) $\Delta\ln(x_t)$; (6) $\Delta^2\ln(x_t)$; (7) $\Delta(x_t/x_{t-1}-1.0)$.

\setcounter{table}{0}   
\renewcommand{\thetable}{A.\arabic{table}}

\begin{center}
\begin{longtable}{ccccc}
\caption{Data description} 
         \label{tab:varidescribe}\\
\toprule
id & tcode & variable      & description                                & group \\
1  & 5     & rsc           & Total Retail Sales of Consumer Goods       & CON   \\
2  & 2     & cfds          & Consumer Sentiment Index                   & CON   \\
3  & 2     & ai            & Amount of Imports                          & IER   \\
4  & 1     & avex          & Average Exchange rate                      & IER   \\
5  & 2     & ae            & Amount of Exports                          & IER   \\
6  & 2     & ceu           & China:Exports: U.S.                         & IER   \\
7  & 2     & aie           & Amount of Imports and Exports              & IER   \\
8  & 1     & tdr3m         & 3-Month Time Deposit Rate                  & IER   \\
9  & 1     & tdr6m         & 6-Month Time Deposit Rate                  & IER   \\
10 & 1     & tdr1y         & 1-Year Time Deposit Rate                   & IER   \\
11 & 1     & tdr3y         & 3-Year Time Deposit Rate                   & IER   \\
12 & 1     & str6m         & 6-Month Short-term Loan Rates              & IER   \\
13 & 1     & str1y         & 1-Year Short-term Loan Rates               & IER   \\
14 & 3     & fic           & Cumulative value of fixed asset investment & INV   \\
15 & 5     & loanml        & Balance of medium- and long-term loans     & MC    \\
16 & 5     & loanba        & Balance of Various Loan                    & MC    \\
17 & 5     & depoba        & Balance of Various Deposits                & MC    \\
18 & 2     & m0            & Narrow Money                               & MC    \\
19 & 2     & m1            & Broad Money                                & MC    \\
20 & 2     & m2            & Extensive M1 Money                         & MC    \\
21 & 7     & rdi           & Total Reserves of Depository Institutions  & MC    \\
22 & 2     & vaix          & Value Added by Industry                    & OUT   \\
23 & 2     & cpitotal      & CPI:Total                                 & PR    \\
24 & 2     & cpifood       & CPI:Food                                  & PR    \\
25 & 2     & cpiapparel    & CPI: Apparel                              & PR    \\
26 & 2     & cpiservice    & CPI: Services                             & PR    \\
27 & 2     & cpimedical    & CPI: Medical Care                         & PR    \\
28 & 2     & cpitransport  & CPI: Transportation                       & PR    \\
29 & 2     & cpiedu        & CPI: Education Culture and Recreation     & PR    \\
30 & 2     & cpiresi       & CPI: Residence                            & PR    \\
31 & 2     & ppitotal      & PPI: Total                                 & PR    \\
32 & 2     & ppiproduce    & PPI: Production                            & PR    \\
33 & 2     & ppiliving     & PPI: Living Information                    & PR    \\
34 & 2     & ppiextract    & PPI: Extractive Industries                 & PR    \\
35 & 2     & ppimaterial   & PPI: Materials                             & PR    \\
36 & 2     & ppiprocessing & PPI: Processing Industries                 & PR    \\
37 & 2     & ppifood       & PPI: Food                                  & PR    \\
38 & 2     & ppiapperal    & PPI: Apparel                               & PR    \\
39 & 2     & ppinondurable & PPI: Non-durable Goods                     & PR    \\
40 & 2     & ppidurable    & PPI: Durable Goods                         & PR    \\
41 & 2     & crbindustral  & CRB: Industry                              & PR    \\
42 & 2     & crbmetal      & CRB: Metal                                 & PR    \\
43 & 2     & crbfood       & CRB: Food                                  & PR    \\
44 & 2     & crbtotal      & CRB: Total                                 & PR    \\
45 & 2     & de            & Dividend Payout Ratio                      & SM    \\
46 & 2     & dp            & Dividend Price Ratio                       & SM    \\
47 & 2     & dy            & Dividend Yield                             & SM    \\
48 & 2     & ep            & Earnings Price Ratio                       & SM    \\
49 & 2     & bm            & Book-to-Market Ratio                       & SM    \\
50 & 2     & svr           & Stock Variance                             & SM    \\
51 & 2     & infl          & Inflation                                  & SM    \\
52 & 2     & ntis          & Net Equity Expansion                       & SM    \\
53 & 2     & mtr           & Monthly Turnover                           & SM    \\
54 & 2     & ret           & Monthly Excess Stock Return                & SM    \\
\bottomrule
\end{longtable}
\end{center}

\section{Additional Results}
\setcounter{table}{0} 
\renewcommand{\thetable}{B.\arabic{table}}
\setcounter{figure}{0}   
\renewcommand{\thefigure}{B.\arabic{figure}}

This table presents the eigenvalues of macroeconomic covariance matrices supervised by the target variables of stock volatility and PPI. The results demonstrate that the SSRF method yields more concentrated eigenvalues compared to alternative methods.

\begin{table}[!ht]
\centering
\renewcommand\arraystretch{1.3}
\caption{Eigenvalues of the  covariance matrices using different PCA techniques.}
\label{tab:add_eigen_concentration}
\resizebox{\textwidth}{!}{
\begin{threeparttable}
    \begin{tabular}{cccccccccccc}
\hline
     &        &  & \multicolumn{4}{c}{Panel B: Stock Volatility} &  & \multicolumn{4}{c}{Panel D: PPI}    \\ \cline{4-7} \cline{9-12} 
     & PCA    &  & sPCA      & srPCA     & SSPCA-f   & SSPCA-d   &  & sPCA   & srPCA  & SSPCA-f & SSPCA-d \\ \hline
1st  & 0.1498 &  & 0.5363    & 0.1418    & 0.5372    & 0.4431    &  & 0.5512 & 0.1984 & 0.5520  & 0.5217  \\
2nd  & 0.1015 &  & 0.1057    & 0.1039    & 0.1058    & 0.0840    &  & 0.1460 & 0.1281 & 0.1461  & 0.1294  \\
3rd  & 0.0816 &  & 0.0677    & 0.0929    & 0.0677    & 0.0723    &  & 0.0582 & 0.0957 & 0.0583  & 0.0618  \\
4th  & 0.0736 &  & 0.0491    & 0.0705    & 0.0492    & 0.0452    &  & 0.0476 & 0.0897 & 0.0476  & 0.0479  \\
5th  & 0.0566 &  & 0.0389    & 0.0558    & 0.0389    & 0.0420    &  & 0.0347 & 0.0594 & 0.0347  & 0.0461  \\
6th  & 0.0477 &  & 0.0290    & 0.0507    & 0.0290    & 0.0389    &  & 0.0324 & 0.0482 & 0.0324  & 0.0318  \\
7th  & 0.0397 &  & 0.0260    & 0.0435    & 0.0257    & 0.0324    &  & 0.0244 & 0.0405 & 0.0244  & 0.0263  \\
8th  & 0.0384 &  & 0.0232    & 0.0416    & 0.0233    & 0.0292    &  & 0.0196 & 0.0355 & 0.0196  & 0.0196  \\
9th  & 0.0332 &  & 0.0183    & 0.0392    & 0.0182    & 0.0289    &  & 0.0158 & 0.0318 & 0.0157  & 0.0155  \\
10th & 0.0322 &  & 0.0176    & 0.0351    & 0.0176    & 0.0272    &  & 0.0136 & 0.0290 & 0.0136  & 0.0143  \\
11th & 0.0299 &  & 0.0161    & 0.0327    & 0.0161    & 0.0210    &  & 0.0097 & 0.0265 & 0.0096  & 0.0105  \\
12th & 0.0286 &  & 0.0109    & 0.0261    & 0.0109    & 0.0193    &  & 0.0067 & 0.0255 & 0.0067  & 0.0093  \\
13th & 0.0241 &  & 0.0090    & 0.0245    & 0.0089    & 0.0156    &  & 0.0054 & 0.0212 & 0.0054  & 0.0084  \\
14th & 0.023  &  & 0.0083    & 0.0234    & 0.0083    & 0.0144    &  & 0.0043 & 0.0192 & 0.0042  & 0.0069  \\
15th & 0.021  &  & 0.0071    & 0.0214    & 0.0071    & 0.0119    &  & 0.0040 & 0.0181 & 0.0040  & 0.0065  \\ \hline
\end{tabular}%
\begin{tablenotes}
\item This table presents the eigenvalues of the in-sample covariance matrices, sorted in descending order and normalized to sum to one.  We compare the factor extraction results of SSPCA-f and SSPCA-d with PCA, sPCA and srPCA. The table is organized into two panels, corresponding to 1-month ahead change in stock volatility (Panel B) and change in PPI (Panel D),  respectively. We retain the top quartile of predictors to represent the threshold parameter $\kappa_1$ and $\kappa_2$ in the screening step.  The sample period is from November 1996 to June 2013.
\end{tablenotes}
\end{threeparttable}%
}

\end{table}
\end{document}